\PassOptionsToPackage{hyphens,spaces,obeyspaces}{url}
\pdfoutput=1
\documentclass[%
 reprint,
 superscriptaddress,
 preprintnumbers,
 nofootinbib,
 amsmath,amssymb,
 aps,
 prl,
]{revtex4-1}

\usepackage{psfrag,slashed,cancel,array,graphicx,todonotes,hyperref}
\usepackage{subcaption}
\usepackage[utf8]{inputenc}
\usepackage{mathtools}
\usepackage{mathrsfs}
\usepackage{multirow}
\usepackage{booktabs} 
\usepackage[normalem]{ulem}

\AtBeginDocument{
\heavyrulewidth=.08em
\lightrulewidth=.05em
\cmidrulewidth=.03em
\belowrulesep=.65ex
\belowbottomsep=0pt
\aboverulesep=.4ex
\abovetopsep=0pt
\cmidrulesep=\doublerulesep
\cmidrulekern=.5em
\defaultaddspace=.5em
}

\usepackage{xkeyval}

\usepackage{tikz,fp}
\usetikzlibrary{external,fixedpointarithmetic,decorations.pathmorphing,positioning,calc,fadings,decorations.markings,decorations.pathreplacing,patterns,arrows}

\tikzset{
  label distance=-3pt,
  >=stealth,
  inner sep=1.5pt,
  boson/.style={
    decoration={snake, segment length=2mm, amplitude=0.5mm},
    decorate,
  },
  quark/.style={
    postaction={decorate},
    decoration={markings,mark=at position .5 with {\draw[->] (-0.01,0) -- (0.07,0);}}
  },
  aquark/.style={
    postaction={decorate},
    decoration={markings,mark=at position .5 with {\draw[->] (0.01,0) -- (-0.07,0);}}
  },
  gluonOld/.style={
    line width=0.7pt,
    decorate,
    decoration={coil,aspect=-0.5,amplitude=-2pt,segment length=2.2pt}
  },
  gluon/.style={
    line width=0.5pt,
    decorate,
    decoration={coil,aspect=-0.75,amplitude=-1.5pt,segment length=2pt}
  },
  agluon/.style={
    line width=0.5pt,
    decorate,
    decoration={coil,aspect=0.75,amplitude=1.5pt,segment length=2pt}
  },
  scalar/.style={
    dashed
  },
  dscalar/.style={
    dashed,
    postaction={decorate},
    decoration={markings,mark=at position 0.65 with {\draw[->] (-0.01,0) -- (0.07,0);}}
  },
  adscalar/.style={
    dashed,
    postaction={decorate},
    decoration={markings,mark=at position 0.65 with {\draw[->] (-0.01,0) -- (0.07,0);}}
  },
  line/.style={
  },
  doubleline/.style={
    double
  },
  middleArrow/.style={
    draw=white,
    decoration={markings,mark=at position 0.65 with{\arrow[scale=1,black]{>}}},
    decorate
  },
  dot/.style={
  	postaction={decorate},
    decoration={markings,mark=between positions 0.5 and 0.5 step 1 with {\draw[fill] (0,0) circle [radius=1.5pt];}}  
  },
  dots/.style={
  	postaction={decorate},
    decoration={markings,mark=between positions 0.3 and 0.6 step 0.3 with {\draw[fill] (0,0) circle [radius=1.5pt];}}  
  },
  dotss/.style={
  	postaction={decorate},
    decoration={markings,mark=between positions 0.25 and 0.75 step 0.25 with {\draw[fill] (0,0) circle [radius=1.5pt];}}  
  },
  blob/.style={
  	pattern=north west lines,
    draw=black
  }
}

\makeatletter
\def\gSetStdExtTypes#1{\presetkeys{gTypes}{eA=#1,eB=#1,eC=#1,eD=#1,eE=#1}{}}
\def\gSetStdIntTypes#1{\presetkeys{gTypes}{iA=#1,iB=#1,iC=#1,iD=#1,iE=#1,iF=#1,iG=#1,iH=#1,iI=#1,iJ=#1}{}}
\def\gSetStdTypes#1{\gSetStdExtTypes{#1}\gSetStdIntTypes{#1}}

\define@cmdkeys{general}{scale}

\define@cmdkeys{general}{rotate}
\define@cmdkeys{general}{font}
\define@cmdkeys{general}{yshift}
\presetkeys{general}{scale=1, rotate=0, font=\scriptsize, yshift=0pt}{}

\define@cmdkeys{gTypes}{eA,eB,eC,eD,eE,iA,iB,iC,iD,iE,iF,iG,iH,iI,iJ}
\gSetStdTypes{gluon}
\define@key{pre}{all}{\gSetStdTypes{#1}}
\define@key{pre}{allE}{\gSetStdExtTypes{#1}}
\define@key{pre}{allI}{\gSetStdIntTypes{#1}}

\define@cmdkeys{gLabels}{eLA,eLB,eLC,eLD,eLE,iLA,iLB,iLC,iLD,iLE,iLF,iLG,iLH,iLI,iLJ}
\presetkeys{gLabels}{eLA={},eLB={},eLC={},eLD={},eLE={},iLA={},iLB={},iLC={},iLD={},iLE={},iLF={},iLG={},iLH={},iLI={},iLJ={}}{}

\def\gTreeTri{\@ifnextchar[\@gTreeTri{\@gTreeTri[]}}
\def\@gTreeTri[#1]#2{%
  \setkeys*{pre}{#1}%
  \setrmkeys{general,gTypes,gLabels}%
  \begin{tikzpicture}
    [line width=1pt,
    baseline={([yshift=-0.5ex+\cmdKV@general@yshift]current bounding box.center)},
    scale=\cmdKV@general@scale,
    rotate=\cmdKV@general@rotate,
    font=\cmdKV@general@font]
    \draw[\cmdKV@gTypes@eA] (0.7,0.3) -- (0.9,0.6) node[label=(\cmdKV@general@rotate):\cmdKV@gLabels@eLA] {};
    \draw[\cmdKV@gTypes@eB] (0.7,0.3) -- (0.9,0) node[label=(\cmdKV@general@rotate):\cmdKV@gLabels@eLB] {};
    \draw[\cmdKV@gTypes@eC] (0.7,0.3) -- (0.4,0.3) node[label=(-180+\cmdKV@general@rotate):\cmdKV@gLabels@eLC] {};
    #2
  \end{tikzpicture}
  \gSetStdTypes{gluon}
}

\def\gTreeS{\@ifnextchar[\@gTreeS{\@gTreeS[]}}
\def\@gTreeS[#1]#2{%
  \setkeys*{pre}{#1}%
  \setrmkeys{general,gTypes,gLabels}%
  \begin{tikzpicture}
    [line width=1pt,
    baseline={([yshift=-0.5ex+\cmdKV@general@yshift]current bounding box.center)},
    scale=\cmdKV@general@scale,
    rotate=\cmdKV@general@rotate,
    font=\cmdKV@general@font]
    \draw[\cmdKV@gTypes@eA] (0.7,0.3) -- (0.9,0.6) node[label=(\cmdKV@general@rotate):\cmdKV@gLabels@eLA] {};
    \draw[\cmdKV@gTypes@eB] (0.7,0.3) -- (0.9,0) node[label=(\cmdKV@general@rotate):\cmdKV@gLabels@eLB] {};
    \draw[\cmdKV@gTypes@eC] (0.2,0.3) -- (0,0) node[label=(-180+\cmdKV@general@rotate):\cmdKV@gLabels@eLC] {};
    \draw[\cmdKV@gTypes@eD] (0.2,0.3) -- (0,0.6) node[label=(180+\cmdKV@general@rotate):\cmdKV@gLabels@eLD] {};
    \draw[label distance=0,\cmdKV@gTypes@iA] (0.7,0.3) -- node[label=(90+\cmdKV@general@rotate):\cmdKV@gLabels@iLA] {} (0.2,0.3);
    #2
  \end{tikzpicture}
  \gSetStdTypes{gluon}
}

\def\gTreeT{\@ifnextchar[\@gTreeT{\@gTreeT[]}}
\def\@gTreeT[#1]#2{%
  \setkeys*{pre}{#1}%
  \setrmkeys{general,gTypes,gLabels}%
  \begin{tikzpicture}
    [line width=1pt,
    baseline={([yshift=-0.5ex+\cmdKV@general@yshift]current bounding box.center)},
    scale=\cmdKV@general@scale,
    rotate=\cmdKV@general@rotate,
    font=\cmdKV@general@font]
    \draw[\cmdKV@gTypes@eA] (0.45,0.5) -- (0.9,0.6) node[label=(\cmdKV@general@rotate):\cmdKV@gLabels@eLA] {};
    \draw[\cmdKV@gTypes@eB] (0.45,0.1) -- (0.9,0) node[label=(\cmdKV@general@rotate):\cmdKV@gLabels@eLB] {};
    \draw[\cmdKV@gTypes@eC] (0.45,0.1) -- (0,0) node[label=(-180+\cmdKV@general@rotate):\cmdKV@gLabels@eLC] {};
    \draw[\cmdKV@gTypes@eD] (0.45,0.5) -- (0,0.6) node[label=(180+\cmdKV@general@rotate):\cmdKV@gLabels@eLD] {};
    \draw[label distance=0,\cmdKV@gTypes@iA] (0.45,0.5) -- node[label=(\cmdKV@general@rotate):\cmdKV@gLabels@iLA] {} (0.45,0.1);
    #2
  \end{tikzpicture}
  \gSetStdTypes{gluon}
}

\def\gTreeU{\@ifnextchar[\@gTreeU{\@gTreeU[]}}
\def\@gTreeU[#1]#2{%
  \setkeys*{pre}{#1}%
  \setrmkeys{general,gTypes,gLabels}%
  \begin{tikzpicture}
    [line width=1pt,
    baseline={([yshift=-0.5ex+\cmdKV@general@yshift]current bounding box.center)},
    scale=\cmdKV@general@scale,
    rotate=\cmdKV@general@rotate,
    font=\cmdKV@general@font]
    \draw[\cmdKV@gTypes@eA] (0.45,0.5) -- (0.9,0.6) node[label=(\cmdKV@general@rotate):\cmdKV@gLabels@eLA] {};
    \draw[\cmdKV@gTypes@eB] (0.45,0.1) -- (0.9,0) node[label=(\cmdKV@general@rotate):\cmdKV@gLabels@eLB] {};
    \draw[\cmdKV@gTypes@eC] (0.45,0.5) -- (0,0) node[label=(-180+\cmdKV@general@rotate):\cmdKV@gLabels@eLC] {};
    \draw[\cmdKV@gTypes@eD] (0.45,0.1) -- (0,0.6) node[label=(180+\cmdKV@general@rotate):\cmdKV@gLabels@eLD] {};
    \draw[label distance=0,\cmdKV@gTypes@iA] (0.45,0.5) -- node[label=(\cmdKV@general@rotate):\cmdKV@gLabels@iLA] {} (0.45,0.1);
    #2
  \end{tikzpicture}
  \gSetStdTypes{gluon}
}

\def\gTadA{\@ifnextchar[\@gTadA{\@gTadA[]}}
\def\@gTadA[#1]#2{%
  \setkeys*{pre}{#1}%
  \setrmkeys{general,gTypes,gLabels}%
  \begin{tikzpicture}
    [line width=1pt,
    baseline={([yshift=-0.5ex+\cmdKV@general@yshift]current bounding box.center)},
    scale=\cmdKV@general@scale,
    rotate=\cmdKV@general@rotate,
    font=\cmdKV@general@font]
    \draw[\cmdKV@gTypes@eA] (0.7,0.3) -- (0.9,0.6) node[label=(\cmdKV@general@rotate):\cmdKV@gLabels@eLA] {};
    \draw[\cmdKV@gTypes@eB] (0.7,0.3) -- (0.9,0) node[label=(\cmdKV@general@rotate):\cmdKV@gLabels@eLB] {};
    \draw[\cmdKV@gTypes@eC] (0.3,0.3) -- (0.2,0) node[label=(-180+\cmdKV@general@rotate):\cmdKV@gLabels@eLC] {};
    \draw[\cmdKV@gTypes@eD] (-0.15,0.45) -- (-0.5,0.6) node[label=(180+\cmdKV@general@rotate):\cmdKV@gLabels@eLD] {};
    \draw[label distance=0,\cmdKV@gTypes@iA] (0.7,0.3) -- node[label=(90+\cmdKV@general@rotate):\cmdKV@gLabels@iLA] {} (0.3,0.3);
    \draw[label distance=0,\cmdKV@gTypes@iB] (0.3,0.3) -- node[label=(-100+\cmdKV@general@rotate):\cmdKV@gLabels@iLB] {} (-0.15,0.45);
    \draw[label distance=0,\cmdKV@gTypes@iC] (-0.15,0.45) -- node[label=(135+\cmdKV@general@rotate):\cmdKV@gLabels@iLC] {} (0.18,0.69);
    \draw[label distance=-0.5,\cmdKV@gTypes@iD] (0.18,0.69) .. controls ($(0.18,0.69) + (-0.12,0.12)$) and ($(0.35,0.87) + (-0.12,0.12)$) .. (0.35,0.87) node[label=(\cmdKV@general@rotate):\cmdKV@gLabels@iLD] {} .. controls ($(0.35,0.87) + (0.12,-0.12)$) and ($(0.18,0.69) + (0.12,-0.12)$) .. (0.18,0.69);
    #2
  \end{tikzpicture}
  \gSetStdTypes{gluon}
}

\def\gTadB{\@ifnextchar[\@gTadB{\@gTadB[]}}
\def\@gTadB[#1]#2{%
  \setkeys*{pre}{#1}%
  \setrmkeys{general,gTypes,gLabels}%
  \begin{tikzpicture}
    [line width=1pt,
    baseline={([yshift=-0.5ex+\cmdKV@general@yshift]current bounding box.center)},
    scale=\cmdKV@general@scale,
    rotate=\cmdKV@general@rotate,
    font=\cmdKV@general@font]
    \draw[\cmdKV@gTypes@eA] (0.7,0.3) -- (0.9,0.6) node[label=(\cmdKV@general@rotate):\cmdKV@gLabels@eLA] {};
    \draw[\cmdKV@gTypes@eB] (0.7,0.3) -- (0.9,0) node[label=(\cmdKV@general@rotate):\cmdKV@gLabels@eLB] {};
    \draw[\cmdKV@gTypes@eC] (0,0.3) -- (-0.2,0) node[label=(-180+\cmdKV@general@rotate):\cmdKV@gLabels@eLC] {};
    \draw[\cmdKV@gTypes@eD] (0,0.3) -- (-0.2,0.6) node[label=(180+\cmdKV@general@rotate):\cmdKV@gLabels@eLD] {};
    \draw[label distance=0,\cmdKV@gTypes@iA] (0.7,0.3) -- node[label=(-90+\cmdKV@general@rotate):\cmdKV@gLabels@iLA] {} (0.35,0.3);
    \draw[label distance=0,\cmdKV@gTypes@iB] (0.35,0.3) -- node[label=(\cmdKV@general@rotate):\cmdKV@gLabels@iLB] {} (0.35,0.6);
    \draw[label distance=-0.5,\cmdKV@gTypes@iC] (0.35,0.6) .. controls (0.18,0.6) and (0.18,0.85) .. (0.35,0.85) .. controls (0.52,0.85) and (0.52,0.6) .. node[label=(\cmdKV@general@rotate):\cmdKV@gLabels@iLC] {} (0.35,0.6);
    \draw[label distance=0,\cmdKV@gTypes@iD] (0.35,0.3) -- node[label=(-90+\cmdKV@general@rotate):\cmdKV@gLabels@iLD] {} (0,0.3);
    #2
  \end{tikzpicture}
  \gSetStdTypes{gluon}
}
\def\gBubA{\@ifnextchar[\@gBubA{\@gBubA[]}}
\def\@gBubA[#1]#2{%
  \setkeys*{pre}{#1}%
  \setrmkeys{general,gTypes,gLabels}%
  \begin{tikzpicture}
    [line width=1pt,
    baseline={([yshift=-0.5ex+\cmdKV@general@yshift]current bounding box.center)},
    scale=\cmdKV@general@scale,
    rotate=\cmdKV@general@rotate,
    font=\cmdKV@general@font]
    \draw[\cmdKV@gTypes@eA] (0.9,0.7) -- (1.2,0.7) node[label=(\cmdKV@general@rotate):\cmdKV@gLabels@eLA] {};
    \draw[\cmdKV@gTypes@eB] (0.5,0.7) -- (0.2,0.7) node[label=(180+\cmdKV@general@rotate):\cmdKV@gLabels@eLB] {};
    \draw[label distance=0,\cmdKV@gTypes@iA] (0.9,0.7) .. controls (0.9,0.45) and (0.5,0.45) .. node[label=(-90+\cmdKV@general@rotate):\cmdKV@gLabels@iLA] {} (0.5,0.7);
    \draw[label distance=0,\cmdKV@gTypes@iB] (0.5,0.7) .. controls (0.5,0.95) and (0.9,0.95) .. node[label=(90+\cmdKV@general@rotate):\cmdKV@gLabels@iLB] {} (0.9,0.7);
    #2
  \end{tikzpicture}
  \gSetStdTypes{gluon}
}
\def\gBubB{\@ifnextchar[\@gBubB{\@gBubB[]}}
\def\@gBubB[#1]#2{%
  \setkeys*{pre}{#1}%
  \setrmkeys{general,gTypes,gLabels}%
  \begin{tikzpicture}
    [line width=1pt,
    baseline={([yshift=-0.5ex+\cmdKV@general@yshift]current bounding box.center)},
    scale=\cmdKV@general@scale,
    rotate=\cmdKV@general@rotate,
    font=\cmdKV@general@font]
    \draw[\cmdKV@gTypes@eA] (1.2,0.7) -- (1.4,1) node[label=(\cmdKV@general@rotate):\cmdKV@gLabels@eLA] {};
    \draw[\cmdKV@gTypes@eB] (1.2,0.7) -- (1.4,0.4) node[label=(\cmdKV@general@rotate):\cmdKV@gLabels@eLB] {};
    \draw[\cmdKV@gTypes@eC] (0.2,0.7) -- (0,0.4) node[label=(-180+\cmdKV@general@rotate):\cmdKV@gLabels@eLC] {};
    \draw[\cmdKV@gTypes@eD] (0.2,0.7) -- (0,1) node[label=(180+\cmdKV@general@rotate):\cmdKV@gLabels@eLD] {};
    \draw[label distance=0,\cmdKV@gTypes@iA] (1.2,0.7) -- node[label=(90+\cmdKV@general@rotate):\cmdKV@gLabels@iLA] {} (0.9,0.7);
    \draw[label distance=0,\cmdKV@gTypes@iB] (0.9,0.7) .. controls (0.9,0.45) and (0.5,0.45) .. node[label=(-90+\cmdKV@general@rotate):\cmdKV@gLabels@iLB] {} (0.5,0.7);
    \draw[label distance=0,\cmdKV@gTypes@iC] (0.5,0.7) .. controls (0.5,0.95) and (0.9,0.95) .. node[label=(90+\cmdKV@general@rotate):\cmdKV@gLabels@iLC] {} (0.9,0.7);
    \draw[label distance=0,\cmdKV@gTypes@iD] (0.5,0.7) -- node[label=(90+\cmdKV@general@rotate):\cmdKV@gLabels@iLD] {} (0.2,0.7);
    #2
  \end{tikzpicture}
  \gSetStdTypes{gluon}
}

\def\gBubC{\@ifnextchar[\@gBubC{\@gBubC[]}}
\def\@gBubC[#1]#2{%
  \setkeys*{pre}{#1}%
  \setrmkeys{general,gTypes,gLabels}%
  \begin{tikzpicture}
    [line width=1pt,
    baseline={([yshift=-0.5ex+\cmdKV@general@yshift]current bounding box.center)},
    scale=\cmdKV@general@scale,
    rotate=\cmdKV@general@rotate,
    font=\cmdKV@general@font]
    \draw[\cmdKV@gTypes@eA] (0.7,0.3) -- (0.9,0.6) node[label=(\cmdKV@general@rotate):\cmdKV@gLabels@eLA] {};
    \draw[\cmdKV@gTypes@eB] (0.7,0.3) -- (0.9,0) node[label=(\cmdKV@general@rotate):\cmdKV@gLabels@eLB] {};
    \draw[\cmdKV@gTypes@eC] (0.3,0.3) -- (0.2,0) node[label=(-180+\cmdKV@general@rotate):\cmdKV@gLabels@eLC] {};
    \draw[\cmdKV@gTypes@eD] (-0.3,0.5) -- (-0.5,0.6) node[label=(180+\cmdKV@general@rotate):\cmdKV@gLabels@eLD] {};
    \draw[label distance=0,\cmdKV@gTypes@iA] (0.7,0.3) -- node[label=(90+\cmdKV@general@rotate):\cmdKV@gLabels@iLA] {} (0.3,0.3);
    \draw[label distance=0,\cmdKV@gTypes@iB] (0.3,0.3) -- node[label=(80+\cmdKV@general@rotate):\cmdKV@gLabels@iLB] {} (0,0.4);
    \draw[label distance=0,\cmdKV@gTypes@iC] (0,0.4) .. controls (-0.08,0.21) and (-0.38,0.3) .. node[label=(-90+\cmdKV@general@rotate):\cmdKV@gLabels@iLC] {} (-0.3,0.5);
    \draw[label distance=0,\cmdKV@gTypes@iD] (-0.3,0.5) .. controls (-0.22,0.65) and (0.03,0.6) .. node[label=(90+\cmdKV@general@rotate):\cmdKV@gLabels@iLD] {} (0,0.4);
    #2
  \end{tikzpicture}
  \gSetStdTypes{gluon}
}
\def\gTriA{\@ifnextchar[\@gTriA{\@gTriA[]}}
\def\@gTriA[#1]#2{%
  \setkeys*{pre}{#1}%
  \setrmkeys{general,gTypes,gLabels}%
  \begin{tikzpicture}
    [line width=1pt,
    baseline={([yshift=-0.5ex+\cmdKV@general@yshift]current bounding box.center)},
    scale=\cmdKV@general@scale,
    rotate=\cmdKV@general@rotate,
    font=\cmdKV@general@font]
    \draw[\cmdKV@gTypes@eA] (0:0.3) -- (0:0.55) node[label=(\cmdKV@general@rotate):\cmdKV@gLabels@eLA] {};
    \draw[\cmdKV@gTypes@eB] (-120:0.3) -- (-120:0.55) node[label=(-180+\cmdKV@general@rotate):\cmdKV@gLabels@eLB] {};
    \draw[\cmdKV@gTypes@eC] (120:0.3) -- (120:0.55) node[label=(180+\cmdKV@general@rotate):\cmdKV@gLabels@eLC] {};
    \draw[label distance=0,\cmdKV@gTypes@iA] (0:0.3) -- node[label=(-60+\cmdKV@general@rotate):\cmdKV@gLabels@iLA] {} (-120:0.3);
    \draw[label distance=0,\cmdKV@gTypes@iB] (-120:0.3) -- node[label=(180+\cmdKV@general@rotate):\cmdKV@gLabels@iLB] {} (120:0.3);
    \draw[label distance=0,\cmdKV@gTypes@iC] (120:0.3) -- node[label=(60+\cmdKV@general@rotate):\cmdKV@gLabels@iLC] {} (0:0.3);
    #2
  \end{tikzpicture}
  \gSetStdTypes{gluon}
}
\def\gTriC{\@ifnextchar[\@gTriC{\@gTriC[]}}
\def\@gTriC[#1]#2{%
  \setkeys*{pre}{#1}%
  \setrmkeys{general,gTypes,gLabels}%
  \begin{tikzpicture}
    [line width=1pt,
    baseline={([yshift=-0.5ex+\cmdKV@general@yshift]current bounding box.center)},
    scale=\cmdKV@general@scale,
    rotate=\cmdKV@general@rotate,
    font=\cmdKV@general@font]
    \draw[\cmdKV@gTypes@eA] (1,0.7) -- (1.3,1.1) node[label=(\cmdKV@general@rotate):\cmdKV@gLabels@eLA] {};
    \draw[\cmdKV@gTypes@eB] (1,0.7) -- (1.3,0.3) node[label=(\cmdKV@general@rotate):\cmdKV@gLabels@eLB] {};
    \draw[\cmdKV@gTypes@eC] (0.3,0.45) -- (0,0.3) node[label=(-180+\cmdKV@general@rotate):\cmdKV@gLabels@eLC] {};
    \draw[\cmdKV@gTypes@eD] (0.3,0.95) -- (0,1.1) node[label=(180+\cmdKV@general@rotate):\cmdKV@gLabels@eLD] {};
    \draw[label distance=0,\cmdKV@gTypes@iA] (1,0.7) -- node[label=(90+\cmdKV@general@rotate):\cmdKV@gLabels@iLA] {} (0.7,0.7);
    \draw[label distance=0,\cmdKV@gTypes@iB] (0.7,0.7) -- node[label=(-70+\cmdKV@general@rotate):\cmdKV@gLabels@iLB] {} (0.3,0.45);
    \draw[label distance=0,\cmdKV@gTypes@iC] (0.3,0.45) -- node[label=(180+\cmdKV@general@rotate):\cmdKV@gLabels@iLC] {} (0.3,0.95);
    \draw[label distance=0,\cmdKV@gTypes@iD] (0.3,0.95) -- node[label=(70+\cmdKV@general@rotate):\cmdKV@gLabels@iLD] {} (0.7,0.7);
    #2
  \end{tikzpicture}
  \gSetStdTypes{gluon}
}
\def\gBox{\@ifnextchar[\@gBox{\@gBox[]}}
\def\@gBox[#1]#2{%
  \setkeys*{pre}{#1}%
  \setrmkeys{general,gTypes,gLabels}%
  \begin{tikzpicture}
    [line width=1pt,
    baseline={([yshift=-0.5ex+\cmdKV@general@yshift]current bounding box.center)},
    scale=\cmdKV@general@scale,
    rotate=\cmdKV@general@rotate,
    font=\cmdKV@general@font]
    \draw[\cmdKV@gTypes@eA] (0.75,0.75) -- (1,1) node[label=(\cmdKV@general@rotate):\cmdKV@gLabels@eLA] {};
    \draw[\cmdKV@gTypes@eB] (0.75,0.25) -- (1,0) node[label=(\cmdKV@general@rotate):\cmdKV@gLabels@eLB] {};
    \draw[\cmdKV@gTypes@eC] (0.25,0.25) -- (0,0) node[label=(-180+\cmdKV@general@rotate):\cmdKV@gLabels@eLC] {};
    \draw[\cmdKV@gTypes@eD] (0.25,0.75) -- (0,1) node[label=(180+\cmdKV@general@rotate):\cmdKV@gLabels@eLD] {};
    \draw[label distance=0,\cmdKV@gTypes@iA] (0.75,0.75) -- node[label=(\cmdKV@general@rotate):\cmdKV@gLabels@iLA] {} (0.75,0.25);
    \draw[label distance=0,\cmdKV@gTypes@iB] (0.75,0.25) -- node[label=(-90+\cmdKV@general@rotate):\cmdKV@gLabels@iLB] {} (0.25,0.25);
    \draw[label distance=0,\cmdKV@gTypes@iC] (0.25,0.25) -- node[label=(180+\cmdKV@general@rotate):\cmdKV@gLabels@iLC] {} (0.25,0.75);
    \draw[label distance=0,\cmdKV@gTypes@iD] (0.25,0.75) -- node[label=(90+\cmdKV@general@rotate):\cmdKV@gLabels@iLD] {} (0.75,0.75);
    #2
  \end{tikzpicture}
  \gSetStdTypes{gluon}
}
\def\gPenta{\@ifnextchar[\@gPenta{\@gPenta[]}}
\def\@gPenta[#1]#2{%
  \setkeys*{pre}{#1}%
  \setrmkeys{general,gTypes,gLabels}%
  \begin{tikzpicture}
    [line width=1pt,
    baseline={([yshift=-0.5ex+\cmdKV@general@yshift]current bounding box.center)},
    scale=\cmdKV@general@scale,
    rotate=\cmdKV@general@rotate,
    font=\cmdKV@general@font]
    \draw[\cmdKV@gTypes@eA] (72:0.35) -- (72:0.75) node[label=(72+\cmdKV@general@rotate):\cmdKV@gLabels@eLA] {};
    \draw[\cmdKV@gTypes@eB] (0:0.35) -- (0:0.75) node[label=(\cmdKV@general@rotate):\cmdKV@gLabels@eLB] {};
    \draw[\cmdKV@gTypes@eC] (-72:0.35) -- (-72:0.75) node[label=(-72+\cmdKV@general@rotate):\cmdKV@gLabels@eLC] {};
    \draw[\cmdKV@gTypes@eD] (-144:0.35) -- (-144:0.75) node[label=(-144+\cmdKV@general@rotate):\cmdKV@gLabels@eLD] {};
    \draw[\cmdKV@gTypes@eE] (144:0.35) -- (144:0.75) node[label=(144+\cmdKV@general@rotate):\cmdKV@gLabels@eLE] {};
    \draw[label distance=0,\cmdKV@gTypes@iA] (72:0.35) -- node[label=(36+\cmdKV@general@rotate):\cmdKV@gLabels@iLA] {} (0:0.35);
    \draw[label distance=0,\cmdKV@gTypes@iB] (0:0.35) -- node[label=(-36+\cmdKV@general@rotate):\cmdKV@gLabels@iLB] {} (-72:0.35);
    \draw[label distance=0,\cmdKV@gTypes@iC] (-72:0.35) -- node[label=(-108+\cmdKV@general@rotate):\cmdKV@gLabels@iLC] {} (-144:0.35);
    \draw[label distance=0,\cmdKV@gTypes@iD] (-144:0.35) -- node[label=(180+\cmdKV@general@rotate):\cmdKV@gLabels@iLD] {} (144:0.35);
    \draw[label distance=0,\cmdKV@gTypes@iE] (144:0.35) -- node[label=(108+\cmdKV@general@rotate):\cmdKV@gLabels@iLE] {} (72:0.35);
    #2
  \end{tikzpicture}
  \gSetStdTypes{gluon}
}

\def\gBoxBox{\@ifnextchar[\@gBoxBox{\@gBoxBox[]}}
\def\@gBoxBox[#1]#2{%
  \setkeys*{pre}{#1}%
  \setrmkeys{general,gTypes,gLabels}%
  \begin{tikzpicture}
    [line width=1pt,
    baseline={([yshift=-0.5ex+\cmdKV@general@yshift]current bounding box.center)},
    scale=\cmdKV@general@scale,
    rotate=\cmdKV@general@rotate,
    font=\cmdKV@general@font]
    \draw[\cmdKV@gTypes@eA] (1.5,0.9) -- (1.8,1.2) node[label=(\cmdKV@general@rotate):\cmdKV@gLabels@eLA] {};
    \draw[\cmdKV@gTypes@eB] (1.5,0.3) -- (1.8,0) node[label=(\cmdKV@general@rotate):\cmdKV@gLabels@eLB] {};
    \draw[\cmdKV@gTypes@eC] (0.3,0.3) -- (0,0) node[label=(-180+\cmdKV@general@rotate):\cmdKV@gLabels@eLC] {};
    \draw[\cmdKV@gTypes@eD] (0.3,0.9) -- (0,1.2) node[label=(180+\cmdKV@general@rotate):\cmdKV@gLabels@eLD] {};
    \draw[label distance=0,\cmdKV@gTypes@iA] (1.5,0.9) -- node[label=(\cmdKV@general@rotate):\cmdKV@gLabels@iLA] {} (1.5,0.3);
    \draw[label distance=0,\cmdKV@gTypes@iB] (1.5,0.3) -- node[label=(-90+\cmdKV@general@rotate):\cmdKV@gLabels@iLB] {} (0.9,0.3);
    \draw[label distance=0,\cmdKV@gTypes@iC] (0.9,0.3) -- node[label=(-90+\cmdKV@general@rotate):\cmdKV@gLabels@iLC] {} (0.3,0.3);
    \draw[label distance=0,\cmdKV@gTypes@iD] (0.3,0.3) -- node[label=(180+\cmdKV@general@rotate):\cmdKV@gLabels@iLD] {} (0.3,0.9);
    \draw[label distance=0,\cmdKV@gTypes@iE] (0.3,0.9) -- node[label=(90+\cmdKV@general@rotate):\cmdKV@gLabels@iLE] {} (0.9,0.9);
    \draw[label distance=0,\cmdKV@gTypes@iF] (0.9,0.9) -- node[label=(90+\cmdKV@general@rotate):\cmdKV@gLabels@iLF] {} (1.5,0.9);
    \draw[label distance=0,\cmdKV@gTypes@iG] (0.9,0.9) -- node[label=(\cmdKV@general@rotate):\cmdKV@gLabels@iLG] {} (0.9,0.3);
    #2
  \end{tikzpicture}
  \gSetStdTypes{gluon}
}
\def\gBoxBoxNP{\@ifnextchar[\@gBoxBoxNP{\@gBoxBoxNP[]}}
\def\@gBoxBoxNP[#1]#2{%
  \setkeys*{pre}{#1}%
  \setrmkeys{general,gTypes,gLabels}%
  \begin{tikzpicture}
    [line width=1pt,
    baseline={([yshift=-0.5ex+\cmdKV@general@yshift]current bounding box.center)},
    scale=\cmdKV@general@scale,
    rotate=\cmdKV@general@rotate,
    font=\cmdKV@general@font]
    \draw[\cmdKV@gTypes@eA] (1.9,1.1) -- (2.2,1.4) node[label=(\cmdKV@general@rotate):\cmdKV@gLabels@eLA] {};
    \draw[\cmdKV@gTypes@eB] (1.9,0.3) -- (2.2,0) node[label=(\cmdKV@general@rotate):\cmdKV@gLabels@eLB] {};
    \draw[\cmdKV@gTypes@eC] (0.3,0.7) -- (0,0.7) node[label=(180+\cmdKV@general@rotate):\cmdKV@gLabels@eLC] {};
    \draw[\cmdKV@gTypes@eD] (1.1,0.7) -- (1.4,0.7) node[label=(\cmdKV@general@rotate):\cmdKV@gLabels@eLD] {};
    \draw[label distance=0,\cmdKV@gTypes@iA] (1.9,1.1) -- node[label=(\cmdKV@general@rotate):\cmdKV@gLabels@iLA] {} (1.9,0.3);
    \draw[label distance=0,\cmdKV@gTypes@iB] (1.9,0.3) -- node[label=(-90+\cmdKV@general@rotate):\cmdKV@gLabels@iLB] {} (0.7,0.3);
    \draw[label distance=0,\cmdKV@gTypes@iC] (0.7,0.3) -- node[label=(-135+\cmdKV@general@rotate):\cmdKV@gLabels@iLC] {} (0.3,0.7);
    \draw[label distance=0,\cmdKV@gTypes@iD] (0.3,0.7) -- node[label=(135+\cmdKV@general@rotate):\cmdKV@gLabels@iLD] {} (0.7,1.1);
    \draw[label distance=0,\cmdKV@gTypes@iE] (0.7,1.1) -- node[label=(90+\cmdKV@general@rotate):\cmdKV@gLabels@iLE] {} (1.9,1.1);
    \draw[label distance=0,\cmdKV@gTypes@iF] (0.7,1.1) -- node[label=(45+\cmdKV@general@rotate):\cmdKV@gLabels@iLF] {} (1.1,0.7);
    \draw[label distance=0,\cmdKV@gTypes@iG] (1.1,0.7) -- node[label=(-45+\cmdKV@general@rotate):\cmdKV@gLabels@iLG] {} (0.7,0.3);
    #2
  \end{tikzpicture}
  \gSetStdTypes{gluon}
}
\def\gTriPenta{\@ifnextchar[\@gTriPenta{\@gTriPenta[]}}
\def\@gTriPenta[#1]#2{%
  \setkeys*{pre}{#1}%
  \setrmkeys{general,gTypes,gLabels}%
  \begin{tikzpicture}
    [line width=1pt,
    baseline={([yshift=-0.5ex+\cmdKV@general@yshift]current bounding box.center)},
    scale=\cmdKV@general@scale,
    rotate=\cmdKV@general@rotate,
    font=\cmdKV@general@font]
    \draw[\cmdKV@gTypes@eA] (72:0.5) -- (45:1) node[label=(35+\cmdKV@general@rotate):\cmdKV@gLabels@eLA] {};
    \draw[\cmdKV@gTypes@eB] (0:0.5) -- (0:1) node[label=(\cmdKV@general@rotate):\cmdKV@gLabels@eLB] {};
    \draw[\cmdKV@gTypes@eC] (-72:0.5) -- (-45:1) node[label=(-35+\cmdKV@general@rotate):\cmdKV@gLabels@eLC] {};
    \draw[\cmdKV@gTypes@eD] (-180:1.2) -- (-180:1.5) node[label=(180+\cmdKV@general@rotate):\cmdKV@gLabels@eLD] {};
    \draw[label distance=0,\cmdKV@gTypes@iA] (72:0.5) -- node[label=(45+\cmdKV@general@rotate):\cmdKV@gLabels@iLA] {} (0:0.5);
    \draw[label distance=0,\cmdKV@gTypes@iB] (0:0.5) -- node[label=(-45+\cmdKV@general@rotate):\cmdKV@gLabels@iLB] {} (-72:0.5);
    \draw[label distance=0,\cmdKV@gTypes@iC] (-72:0.5) -- node[label=(-108+\cmdKV@general@rotate):\cmdKV@gLabels@iLC] {} (-144:0.5);
    \draw[label distance=0,\cmdKV@gTypes@iD] (-144:0.5) -- node[label=(-108+\cmdKV@general@rotate):\cmdKV@gLabels@iLD] {} (-180:1.2);
    \draw[label distance=2,\cmdKV@gTypes@iE] (-180:1.2) -- node[label=(90+\cmdKV@general@rotate):\cmdKV@gLabels@iLE] {} (144:0.5);
    \draw[label distance=2,\cmdKV@gTypes@iF] (144:0.5) -- node[label=(90+\cmdKV@general@rotate):\cmdKV@gLabels@iLF] {} (72:0.5);
    \draw[label distance=0,\cmdKV@gTypes@iG] (144:0.5) -- node[label=(\cmdKV@general@rotate):\cmdKV@gLabels@iLG] {} (-144:0.5);
    #2
  \end{tikzpicture}
  \gSetStdTypes{gluon}
}
\def\gTriTri{\@ifnextchar[\@gTriTri{\@gTriTri[]}}
\def\@gTriTri[#1]#2{%
  \setkeys*{pre}{#1}%
  \setrmkeys{general,gTypes,gLabels}%
  \begin{tikzpicture}
    [line width=1pt,
    baseline={([yshift=-0.5ex+\cmdKV@general@yshift]current bounding box.center)},
    scale=\cmdKV@general@scale,
    rotate=\cmdKV@general@rotate,
    font=\cmdKV@general@font]
    \draw[\cmdKV@gTypes@eA] (1.9,1.1) -- (2.2,1.4) node[label=(\cmdKV@general@rotate):\cmdKV@gLabels@eLA] {};
    \draw[\cmdKV@gTypes@eB] (1.9,0.3) -- (2.2,0) node[label=(\cmdKV@general@rotate):\cmdKV@gLabels@eLB] {};
    \draw[\cmdKV@gTypes@eC] (0.3,0.3) -- (0,0) node[label=(-180+\cmdKV@general@rotate):\cmdKV@gLabels@eLC] {};
    \draw[\cmdKV@gTypes@eD] (0.3,1.1) -- (0,1.4) node[label=(180+\cmdKV@general@rotate):\cmdKV@gLabels@eLD] {};
    \draw[label distance=0,\cmdKV@gTypes@iA] (1.9,1.1) -- node[label=(\cmdKV@general@rotate):\cmdKV@gLabels@iLA] {} (1.9,0.3);
    \draw[label distance=0,\cmdKV@gTypes@iB] (1.9,0.3) -- node[label=(-135+\cmdKV@general@rotate):\cmdKV@gLabels@iLB] {} (1.3,0.7);
    \draw[label distance=0,\cmdKV@gTypes@iC] (1.3,0.7) -- node[label=(135+\cmdKV@general@rotate):\cmdKV@gLabels@iLC] {} (1.9,1.1);
    \draw[label distance=0,\cmdKV@gTypes@iD] (1.3,0.7) -- node[label=(90+\cmdKV@general@rotate):\cmdKV@gLabels@iLD] {} (0.9,0.7);
    \draw[label distance=0,\cmdKV@gTypes@iE] (0.9,0.7) -- node[label=(-45+\cmdKV@general@rotate):\cmdKV@gLabels@iLE] {} (0.3,0.3);
    \draw[label distance=0,\cmdKV@gTypes@iF] (0.3,0.3) -- node[label=(180+\cmdKV@general@rotate):\cmdKV@gLabels@iLF] {} (0.3,1.1);
    \draw[label distance=0,\cmdKV@gTypes@iG] (0.3,1.1) -- node[label=(45+\cmdKV@general@rotate):\cmdKV@gLabels@iLG] {} (0.9,0.7);
    #2
  \end{tikzpicture}
  \gSetStdTypes{gluon}
}

\def\gBoxBubA{\@ifnextchar[\@gBoxBubA{\@gBoxBubA[]}}
\def\@gBoxBubA[#1]#2{%
  \setkeys*{pre}{#1}%
  \setrmkeys{general,gTypes,gLabels}%
  \begin{tikzpicture}
    [line width=1pt,
    baseline={([yshift=-0.5ex+\cmdKV@general@yshift]current bounding box.center)},
    scale=\cmdKV@general@scale,
    rotate=\cmdKV@general@rotate,
    font=\cmdKV@general@font]
    \draw[\cmdKV@gTypes@eA] (0.9,0.9) -- (1.1,1.1) node[label=(\cmdKV@general@rotate):\cmdKV@gLabels@eLA] {};
    \draw[\cmdKV@gTypes@eB] (0.9,0.1) -- (1.1,-0.1) node[label=(\cmdKV@general@rotate):\cmdKV@gLabels@eLB] {};
    \draw[\cmdKV@gTypes@eC] (0.1,0.1) -- (-0.1,-0.1) node[label=(-180+\cmdKV@general@rotate):\cmdKV@gLabels@eLC] {};
    \draw[\cmdKV@gTypes@eD] (0.1,0.9) -- (-0.1,1.1) node[label=(180+\cmdKV@general@rotate):\cmdKV@gLabels@eLD] {};
    \draw[label distance=0,\cmdKV@gTypes@iA] (0.9,0.9) -- node[label=(\cmdKV@general@rotate):\cmdKV@gLabels@iLA] {} (0.9,0.1);
    \draw[label distance=0,\cmdKV@gTypes@iB] (0.9,0.1) -- node[label=(-90+\cmdKV@general@rotate):\cmdKV@gLabels@iLB] {} (0.1,0.1);
    \draw[label distance=0,\cmdKV@gTypes@iC] (0.1,0.1) -- node[label=(180+\cmdKV@general@rotate):\cmdKV@gLabels@iLC] {} (0.1,0.9);
    \draw[label distance=0,\cmdKV@gTypes@iD] (0.1,0.9) -- node[label=(90+\cmdKV@general@rotate):\cmdKV@gLabels@iLD] {} (0.32,0.9);
    \draw[label distance=0,\cmdKV@gTypes@iE] (0.32,0.9) .. controls (0.32,1.15) and (0.68,1.15) .. node[label=(90+\cmdKV@general@rotate):\cmdKV@gLabels@iLE] {} (0.68,0.9);
    \draw[label distance=0,\cmdKV@gTypes@iF] (0.68,0.9) .. controls (0.68,0.65) and (0.32,0.65) .. node[label=(-90+\cmdKV@general@rotate):\cmdKV@gLabels@iLF] {} (0.32,0.9);
    \draw[label distance=0,\cmdKV@gTypes@iG] (0.68,0.9) -- node[label=(90+\cmdKV@general@rotate):\cmdKV@gLabels@iLG] {} (0.9,0.9);
    #2
  \end{tikzpicture}
  \gSetStdTypes{gluon}
}

\def\gBoxBubB{\@ifnextchar[\@gBoxBubB{\@gBoxBubB[]}}
\def\@gBoxBubB[#1]#2{%
  \setkeys*{pre}{#1}%
  \setrmkeys{general,gTypes,gLabels}%
  \begin{tikzpicture}
    [line width=1pt,
    baseline={([yshift=-0.5ex+\cmdKV@general@yshift]current bounding box.center)},
    scale=\cmdKV@general@scale,
    rotate=\cmdKV@general@rotate,
    font=\cmdKV@general@font]
    \draw[label distance=-1,\cmdKV@gTypes@eA] (0,0.4) -- (0,0.7) node[label=(\cmdKV@general@rotate):\cmdKV@gLabels@eLA] {};
    \draw[\cmdKV@gTypes@eB] (0.4,0) -- (0.7,0) node[label=(\cmdKV@general@rotate):\cmdKV@gLabels@eLB] {};
    \draw[label distance=-1,\cmdKV@gTypes@eC] (0,-0.4) -- (0,-0.7) node[label=(\cmdKV@general@rotate):\cmdKV@gLabels@eLC] {};
    \draw[\cmdKV@gTypes@eD] (-1.1,0) -- (-1.4,0) node[label=(180+\cmdKV@general@rotate):\cmdKV@gLabels@eLD] {};
    \draw[label distance=0,\cmdKV@gTypes@iA] (0,0.4) -- node[label=(45+\cmdKV@general@rotate):\cmdKV@gLabels@iLA] {} (0.4,0);
    \draw[label distance=0,\cmdKV@gTypes@iB] (0.4,0) -- node[label=(-45+\cmdKV@general@rotate):\cmdKV@gLabels@iLB] {} (0,-0.4);
    \draw[label distance=0,\cmdKV@gTypes@iC] (0,-0.4) -- node[label=(-135+\cmdKV@general@rotate):\cmdKV@gLabels@iLC] {} (-0.4,0);
    \draw[label distance=0,\cmdKV@gTypes@iD] (-0.4,0) -- node[label=(135+\cmdKV@general@rotate):\cmdKV@gLabels@iLD] {} (0,0.4);
    \draw[label distance=0,\cmdKV@gTypes@iE] (-0.4,0) --  node[label=(90+\cmdKV@general@rotate):\cmdKV@gLabels@iLE] {} (-0.7,0);
    \draw[label distance=0,\cmdKV@gTypes@iF] (-0.7,0) .. controls (-0.7,-0.26) and (-1.1,-0.26) .. node[label=(-90+\cmdKV@general@rotate):\cmdKV@gLabels@iLF] {} (-1.1,0);
    \draw[label distance=0,\cmdKV@gTypes@iG] (-1.1,0) .. controls (-1.1,0.26) and (-0.7,0.26) ..  node[label=(90+\cmdKV@general@rotate):\cmdKV@gLabels@iLG] {} (-0.7,0);
    #2
  \end{tikzpicture}
  \gSetStdTypes{gluon}
}

\def\gPentaTad{\@ifnextchar[\@gPentaTad{\@gPentaTad[]}}
\def\@gPentaTad[#1]#2{%
  \setkeys*{pre}{#1}%
  \setrmkeys{general,gTypes,gLabels}%
  \begin{tikzpicture}
    [line width=1pt,
    baseline={([yshift=-0.5ex+\cmdKV@general@yshift]current bounding box.center)},
    scale=\cmdKV@general@scale,
    rotate=\cmdKV@general@rotate,
    font=\cmdKV@general@font]
    \draw[label distance=-0.5,\cmdKV@gTypes@eA] (108:0.35) -- (108:0.75) node[label=(180+\cmdKV@general@rotate):\cmdKV@gLabels@eLA] {};
    \draw[\cmdKV@gTypes@eB] (36:0.35) -- (36:0.75) node[label=(36+\cmdKV@general@rotate):\cmdKV@gLabels@eLB] {};
    \draw[\cmdKV@gTypes@eC] (-36:0.35) -- (-36:0.75) node[label=(-36+\cmdKV@general@rotate):\cmdKV@gLabels@eLC] {};
    \draw[label distance=-0.5,\cmdKV@gTypes@eD] (-108:0.35) -- (-108:0.75) node[label=(-180+\cmdKV@general@rotate):\cmdKV@gLabels@eLD] {};
    \draw[label distance=0,\cmdKV@gTypes@iA] (108:0.35) -- node[label=(72+\cmdKV@general@rotate):\cmdKV@gLabels@iLA] {} (36:0.35);
    \draw[label distance=0,\cmdKV@gTypes@iB] (36:0.35) -- node[label=(\cmdKV@general@rotate):\cmdKV@gLabels@iLB] {} (-36:0.35);
    \draw[label distance=0,\cmdKV@gTypes@iC] (-36:0.35) -- node[label=(-72+\cmdKV@general@rotate):\cmdKV@gLabels@iLC] {} (-108:0.35);
    \draw[label distance=0,\cmdKV@gTypes@iD] (-108:0.35) -- node[label=(-144+\cmdKV@general@rotate):\cmdKV@gLabels@iLD] {} (-180:0.35);
    \draw[label distance=0,\cmdKV@gTypes@iE] (180:0.35) -- node[label=(144+\cmdKV@general@rotate):\cmdKV@gLabels@iLE] {} (108:0.35);
    \draw[label distance=0,\cmdKV@gTypes@iF] (180:0.35) -- node[label=(90+\cmdKV@general@rotate):\cmdKV@gLabels@iLF] {} (180:0.75);
    \draw[label distance=-0.5,\cmdKV@gTypes@iG] (180:0.75) .. controls ($(180:0.75)+(0,-0.18)$) and ($(180:1)+(0,-0.18)$) .. (180:1) node[label=(180+\cmdKV@general@rotate):\cmdKV@gLabels@iLG] {} .. controls ($(180:1)+(0,0.18)$) and ($(180:0.75)+(0,0.18)$) .. (180:0.75);
    #2
  \end{tikzpicture}
  \gSetStdTypes{gluon}
}

\def\gBoxTadA{\@ifnextchar[\@gBoxTadA{\@gBoxTadA[]}}
\def\@gBoxTadA[#1]#2{%
  \setkeys*{pre}{#1}%
  \setrmkeys{general,gTypes,gLabels}%
  \begin{tikzpicture}
    [line width=1pt,
    baseline={([yshift=-0.5ex+\cmdKV@general@yshift]current bounding box.center)},
    scale=\cmdKV@general@scale,
    rotate=\cmdKV@general@rotate,
    font=\cmdKV@general@font]
    \draw[label distance=-0.5,\cmdKV@gTypes@eA] (0,0.4) -- (0,0.7) node[label=(\cmdKV@general@rotate):\cmdKV@gLabels@eLA] {};
    \draw[\cmdKV@gTypes@eB] (0.4,0) -- (0.7,0) node[label=(\cmdKV@general@rotate):\cmdKV@gLabels@eLB] {};
    \draw[label distance=-0.5,\cmdKV@gTypes@eC] (0,-0.4) -- (0,-0.7) node[label=(\cmdKV@general@rotate):\cmdKV@gLabels@eLC] {};
    \draw[\cmdKV@gTypes@eD] (-0.7,0) -- (-0.7,-0.3) node[label=(-90+\cmdKV@general@rotate):\cmdKV@gLabels@eLD] {};
    \draw[label distance=0,\cmdKV@gTypes@iA] (0,0.4) -- node[label=(45+\cmdKV@general@rotate):\cmdKV@gLabels@iLA] {} (0.4,0);
    \draw[label distance=0,\cmdKV@gTypes@iB] (0.4,0) -- node[label=(-45+\cmdKV@general@rotate):\cmdKV@gLabels@iLB] {} (0,-0.4);
    \draw[label distance=0,\cmdKV@gTypes@iC] (0,-0.4) -- node[label=(-135+\cmdKV@general@rotate):\cmdKV@gLabels@iLC] {} (-0.4,0);
    \draw[label distance=0,\cmdKV@gTypes@iD] (-0.4,0) -- node[label=(135+\cmdKV@general@rotate):\cmdKV@gLabels@iLD] {} (0,0.4);
    \draw[label distance=0,\cmdKV@gTypes@iE] (-0.4,0) --  node[label=(90+\cmdKV@general@rotate):\cmdKV@gLabels@iLE] {} (-0.7,0);
    \draw[label distance=0,\cmdKV@gTypes@iF] (-0.7,0) -- node[label=(90+\cmdKV@general@rotate):\cmdKV@gLabels@iLF] {} (-1.1,0);
    \draw[label distance=-0.5,\cmdKV@gTypes@iG] (-1.1,0) .. controls (-1.1,-0.18) and (-1.35,-0.18) .. (-1.35,0) node[label=(180+\cmdKV@general@rotate):\cmdKV@gLabels@iLG] {} .. controls (-1.35,0.18) and (-1.1,0.18) .. (-1.1,0);
    #2
  \end{tikzpicture}
  \gSetStdTypes{gluon}
}

\def\gBoxBoxBoxA{\@ifnextchar[\@gBoxBoxBoxA{\@gBoxBoxBoxA[]}}
\def\@gBoxBoxBoxA[#1]#2{%
  \setkeys*{pre}{#1}%
  \setrmkeys{general,gTypes,gLabels}%
  \begin{tikzpicture}
    [line width=1pt,
    baseline={([yshift=-0.5ex+\cmdKV@general@yshift]current bounding box.center)},
    scale=\cmdKV@general@scale,
    rotate=\cmdKV@general@rotate,
    font=\cmdKV@general@font]
    \draw[\cmdKV@gTypes@eA] (1.5,0.9) -- (1.8,1.2) node[label=(\cmdKV@general@rotate):\cmdKV@gLabels@eLA] {};
    \draw[\cmdKV@gTypes@eB] (1.5,0.3) -- (1.8,0) node[label=(\cmdKV@general@rotate):\cmdKV@gLabels@eLB] {};
    \draw[\cmdKV@gTypes@eC] (-0.3,0.3) -- (-0.6,0) node[label=(-180+\cmdKV@general@rotate):\cmdKV@gLabels@eLC] {};
    \draw[\cmdKV@gTypes@eD] (-0.3,0.9) -- (-0.6,1.2) node[label=(180+\cmdKV@general@rotate):\cmdKV@gLabels@eLD] {};
    \draw[label distance=0,\cmdKV@gTypes@iA] (1.5,0.9) -- node[label=(\cmdKV@general@rotate):\cmdKV@gLabels@iLA] {} (1.5,0.3);
    \draw[label distance=0,\cmdKV@gTypes@iB] (1.5,0.3) -- node[label=(-90+\cmdKV@general@rotate):\cmdKV@gLabels@iLB] {} (0.9,0.3);
    \draw[label distance=0,\cmdKV@gTypes@iC] (0.9,0.3) -- node[label=(-90+\cmdKV@general@rotate):\cmdKV@gLabels@iLC] {} (0.3,0.3);
    \draw[label distance=0,\cmdKV@gTypes@iD] (0.3,0.3) -- node[label=(-90+\cmdKV@general@rotate):\cmdKV@gLabels@iLD] {} (-0.3,0.3);
    \draw[label distance=0,\cmdKV@gTypes@iE] (-0.3,0.3) -- node[label=(180+\cmdKV@general@rotate):\cmdKV@gLabels@iLE] {} (-0.3,0.9);
    \draw[label distance=0,\cmdKV@gTypes@iF] (-0.3,0.9) -- node[label=(90+\cmdKV@general@rotate):\cmdKV@gLabels@iLF] {} (0.3,0.9);
    \draw[label distance=0,\cmdKV@gTypes@iG] (0.3,0.9) -- node[label=(90+\cmdKV@general@rotate):\cmdKV@gLabels@iLG] {} (0.9,0.9);
    \draw[label distance=0,\cmdKV@gTypes@iH] (0.9,0.9) -- node[label=(90+\cmdKV@general@rotate):\cmdKV@gLabels@iLH] {} (1.5,0.9);
    \draw[label distance=0,\cmdKV@gTypes@iI] (0.9,0.9) -- node[label=(\cmdKV@general@rotate):\cmdKV@gLabels@iLI] {} (0.9,0.3);
    \draw[label distance=0,\cmdKV@gTypes@iJ] (0.3,0.9) -- node[label=(\cmdKV@general@rotate):\cmdKV@gLabels@iLJ] {} (0.3,0.3);
    #2
  \end{tikzpicture}
  \gSetStdTypes{gluon}
}

\def\gBoxBoxBoxB{\@ifnextchar[\@gBoxBoxBoxB{\@gBoxBoxBoxB[]}}
\def\@gBoxBoxBoxB[#1]#2{%
  \setkeys*{pre}{#1}%
  \setrmkeys{general,gTypes,gLabels}%
  \begin{tikzpicture}
    [line width=1pt,
    baseline={([yshift=-0.5ex+\cmdKV@general@yshift]current bounding box.center)},
    scale=\cmdKV@general@scale,
    rotate=\cmdKV@general@rotate,
    font=\cmdKV@general@font]
    \draw[\cmdKV@gTypes@eA] (1.5,1.5) -- (1.8,1.8) node[label=(\cmdKV@general@rotate):\cmdKV@gLabels@eLA] {};
    \draw[\cmdKV@gTypes@eB] (1.5,0.3) -- (1.8,0) node[label=(\cmdKV@general@rotate):\cmdKV@gLabels@eLB] {};
    \draw[\cmdKV@gTypes@eC] (0.3,0.3) -- (0,0) node[label=(-180+\cmdKV@general@rotate):\cmdKV@gLabels@eLC] {};
    \draw[\cmdKV@gTypes@eD] (0.3,1.5) -- (0,1.8) node[label=(180+\cmdKV@general@rotate):\cmdKV@gLabels@eLD] {};
    \draw[label distance=0,\cmdKV@gTypes@iA] (1.5,1.5) -- node[label=(\cmdKV@general@rotate):\cmdKV@gLabels@iLA] {} (1.5,0.3);
    \draw[label distance=0,\cmdKV@gTypes@iB] (1.5,0.3) -- node[label=(-90+\cmdKV@general@rotate):\cmdKV@gLabels@iLB] {} (0.9,0.3);
    \draw[label distance=0,\cmdKV@gTypes@iC] (0.9,0.3) -- node[label=(-90+\cmdKV@general@rotate):\cmdKV@gLabels@iLC] {} (0.3,0.3);
    \draw[label distance=0,\cmdKV@gTypes@iD] (0.3,0.3) -- node[label=(180+\cmdKV@general@rotate):\cmdKV@gLabels@iLD] {} (0.3,0.9);
    \draw[label distance=0,\cmdKV@gTypes@iE] (0.3,0.9) -- node[label=(180+\cmdKV@general@rotate):\cmdKV@gLabels@iLE] {} (0.3,1.5);
    \draw[label distance=0,\cmdKV@gTypes@iF] (0.3,1.5) -- node[label=(90+\cmdKV@general@rotate):\cmdKV@gLabels@iLF] {} (0.9,1.5);
    \draw[label distance=0,\cmdKV@gTypes@iG] (0.9,1.5) -- node[label=(90+\cmdKV@general@rotate):\cmdKV@gLabels@iLG] {} (1.5,1.5);
    \draw[label distance=0,\cmdKV@gTypes@iH] (0.9,1.5) -- node[label=(\cmdKV@general@rotate):\cmdKV@gLabels@iLH] {} (0.9,0.9);
    \draw[label distance=0,\cmdKV@gTypes@iI] (0.9,0.9) -- node[label=(\cmdKV@general@rotate):\cmdKV@gLabels@iLI] {} (0.9,0.3);
    \draw[label distance=0,\cmdKV@gTypes@iJ] (0.9,0.9) -- node[label=(90+\cmdKV@general@rotate):\cmdKV@gLabels@iLJ] {} (0.3,0.9);
    #2
  \end{tikzpicture}
  \gSetStdTypes{gluon}
}

\def\cBoxA{\@ifnextchar[\@cBoxA{\@cBoxA[]}}
\def\@cBoxA[#1]#2{%
  \setkeys*{pre}{#1}%
  \setrmkeys{general,gTypes,gLabels}%
  \begin{tikzpicture}
    [line width=1pt,
    baseline={([yshift=-0.5ex+\cmdKV@general@yshift]current bounding box.center)},
    scale=\cmdKV@general@scale,
    rotate=\cmdKV@general@rotate,
    font=\cmdKV@general@font]
    \fill[blob] (0.4,0) ellipse (0.2 and 0.4);
    \fill[blob] (-0.4,0) ellipse (0.2 and 0.4);
    \draw[\cmdKV@gTypes@eA] ($ (0.4,0) + (45:0.26) $) -- ($ (0.4,0) + (45:0.7) $) node[label=(\cmdKV@general@rotate):\cmdKV@gLabels@eLA] {};
    \draw[\cmdKV@gTypes@eB] ($ (0.4,0) + (-45:0.26) $) -- ($ (0.4,0) + (-45:0.7) $) node[label=(\cmdKV@general@rotate):\cmdKV@gLabels@eLB] {};
    \draw[\cmdKV@gTypes@eC] ($ (-0.4,0) + (-135:0.26) $) -- ($ (-0.4,0) + (-135:0.7) $) node[label=(-180+\cmdKV@general@rotate):\cmdKV@gLabels@eLC] {};
    \draw[\cmdKV@gTypes@eD] ($ (-0.4,0) + (135:0.26) $) -- ($ (-0.4,0) + (135:0.7) $) node[label=(180+\cmdKV@general@rotate):\cmdKV@gLabels@eLD] {};
    \draw[label distance=1,\cmdKV@gTypes@iA] ($ (0.4,0) + (-135:0.26) $) -- node[label=(-90+\cmdKV@general@rotate):\cmdKV@gLabels@iLA] {}($ (-0.4,0) + (-45:0.26) $);
    \draw[label distance=1,\cmdKV@gTypes@iB] ($ (-0.4,0) + (45:0.26) $) -- node[label=(90+\cmdKV@general@rotate):\cmdKV@gLabels@iLB] {}($ (0.4,0) + (135:0.26) $);
    \node[label=(\cmdKV@general@rotate):\cmdKV@gLabels@iLC] at (0.6,0) {};
    \node[label=(180+\cmdKV@general@rotate):\cmdKV@gLabels@iLD] at (-0.6,0) {};
    #2
  \end{tikzpicture}
  \gSetStdTypes{gluon}
}

\def\cBoxB{\@ifnextchar[\@cBoxB{\@cBoxB[]}}
\def\@cBoxB[#1]#2{%
  \setkeys*{pre}{#1}%
  \setrmkeys{general,gTypes,gLabels}%
  \begin{tikzpicture}
    [line width=1pt,
    baseline={([yshift=-0.5ex+\cmdKV@general@yshift]current bounding box.center)},
    scale=\cmdKV@general@scale,
    rotate=\cmdKV@general@rotate,
    font=\cmdKV@general@font]
    \fill[blob] (0,0.3) ellipse (0.4 and 0.2);
    \fill[blob] (0,-0.3) ellipse (0.4 and 0.2);
    \draw[\cmdKV@gTypes@eA] ($ (0,0.3) + (0:0.4) $) -- ($ (0,0.3) + (10:0.7) $) node[label=(\cmdKV@general@rotate):\cmdKV@gLabels@eLA] {};
    \draw[\cmdKV@gTypes@eB] ($ (0,-0.3) + (0:0.4) $) -- ($ (0,-0.3) + (-10:0.7) $) node[label=(\cmdKV@general@rotate):\cmdKV@gLabels@eLB] {};
    \draw[\cmdKV@gTypes@eC] ($ (0,-0.3) + (180:0.4) $) -- ($ (0,-0.3) + (-170:0.7) $) node[label=(180+\cmdKV@general@rotate):\cmdKV@gLabels@eLC] {};
    \draw[\cmdKV@gTypes@eD] ($ (0,0.3) + (180:0.4) $) -- ($ (0,0.3) + (170:0.7) $) node[label=(180+\cmdKV@general@rotate):\cmdKV@gLabels@eLD] {};
    \draw[label distance=1,\cmdKV@gTypes@iA] ($ (0,0.3) + (-45:0.26) $) -- node[label=(\cmdKV@general@rotate):\cmdKV@gLabels@iLA] {}($ (0,-0.3) + (45:0.26) $);
    \draw[label distance=1,\cmdKV@gTypes@iA] ($ (0,-0.3) + (135:0.26) $) -- node[label=(-180+\cmdKV@general@rotate):\cmdKV@gLabels@iLB] {}($ (0,0.3) + (-135:0.26) $);
    \node[label=(-90+\cmdKV@general@rotate):\cmdKV@gLabels@iLC] at (0,-0.5) {};
    \node[label=(90+\cmdKV@general@rotate):\cmdKV@gLabels@iLD] at (0,0.5) {};
    #2
  \end{tikzpicture}
  \gSetStdTypes{gluon}
}

\def\cBoxBoxA{\@ifnextchar[\@cBoxBoxA{\@cBoxBoxA[]}}
\def\@cBoxBoxA[#1]#2{%
  \setkeys*{pre}{#1}%
  \setrmkeys{general,gTypes,gLabels}%
  \begin{tikzpicture}
    [line width=1pt,
    baseline={([yshift=-0.5ex+\cmdKV@general@yshift]current bounding box.center)},
    scale=\cmdKV@general@scale,
    rotate=\cmdKV@general@rotate,
    font=\cmdKV@general@font]
    \fill[blob] (0.4,0) ellipse (0.2 and 0.4);
    \fill[blob] (-0.4,0) ellipse (0.2 and 0.4);
    \fill[blob] (1.2,0) ellipse (0.2 and 0.4);
    \draw[\cmdKV@gTypes@eA] ($ (1.2,0) + (45:0.26) $) -- ($ (1.2,0) + (45:0.7) $) node[label=(\cmdKV@general@rotate):\cmdKV@gLabels@eLA] {};
    \draw[\cmdKV@gTypes@eB] ($ (1.2,0) + (-45:0.26) $) -- ($ (1.2,0) + (-45:0.7) $) node[label=(\cmdKV@general@rotate):\cmdKV@gLabels@eLB] {};
    \draw[\cmdKV@gTypes@eC] ($ (-0.4,0) + (-135:0.26) $) -- ($ (-0.4,0) + (-135:0.7) $) node[label=(-180+\cmdKV@general@rotate):\cmdKV@gLabels@eLC] {};
    \draw[\cmdKV@gTypes@eD] ($ (-0.4,0) + (135:0.26) $) -- ($ (-0.4,0) + (135:0.7) $) node[label=(180+\cmdKV@general@rotate):\cmdKV@gLabels@eLD] {};
    \draw[label distance=1,\cmdKV@gTypes@iA] ($ (1.2,0) + (-135:0.26) $) -- node[label=(-90+\cmdKV@general@rotate):\cmdKV@gLabels@iLA] {}($ (0.4,0) + (-45:0.26) $);
    \draw[label distance=1,\cmdKV@gTypes@iB] ($ (0.4,0) + (-135:0.26) $) -- node[label=(-90+\cmdKV@general@rotate):\cmdKV@gLabels@iLB] {}($ (-0.4,0) + (-45:0.26) $);
    \draw[label distance=1,\cmdKV@gTypes@iC] ($ (-0.4,0) + (45:0.26) $) -- node[label=(90+\cmdKV@general@rotate):\cmdKV@gLabels@iLC] {}($ (0.4,0) + (135:0.26) $);
    \draw[label distance=1,\cmdKV@gTypes@iD] ($ (0.4,0) + (45:0.26) $) -- node[label=(90+\cmdKV@general@rotate):\cmdKV@gLabels@iLD] {}($ (1.2,0) + (135:0.26) $);
    \node[label=(\cmdKV@general@rotate):\cmdKV@gLabels@iLE] at (1.4,0) {};
    \node[label=(\cmdKV@general@rotate):\cmdKV@gLabels@iLF] at (0.6,0) {};
    \node[label=(180+\cmdKV@general@rotate):\cmdKV@gLabels@iLG] at (-0.6,0) {};
    #2
  \end{tikzpicture}
  \gSetStdTypes{gluon}
}

\def\cBoxBoxB{\@ifnextchar[\@cBoxBoxB{\@cBoxBoxB[]}}
\def\@cBoxBoxB[#1]#2{%
  \setkeys*{pre}{#1}%
  \setrmkeys{general,gTypes,gLabels}%
  \begin{tikzpicture}
    [line width=1pt,
    baseline={([yshift=-0.5ex+\cmdKV@general@yshift]current bounding box.center)},
    scale=\cmdKV@general@scale,
    rotate=\cmdKV@general@rotate,
    font=\cmdKV@general@font]
    \fill[blob] (0,0.3) ellipse (0.4 and 0.2);
    \fill[blob] (0,-0.3) ellipse (0.4 and 0.2);
    \fill[blob] (1,0) ellipse (0.2 and 0.4);
    \draw[\cmdKV@gTypes@eA] ($ (1,0) + (45:0.26) $) -- ($ (1,0) + (45:0.7) $) node[label=(\cmdKV@general@rotate):\cmdKV@gLabels@eLA] {};
    \draw[\cmdKV@gTypes@eB] ($ (1,0) + (-45:0.26) $) -- ($ (1,0) + (-45:0.7) $) node[label=(\cmdKV@general@rotate):\cmdKV@gLabels@eLB] {};
    \draw[\cmdKV@gTypes@eC] ($ (0,-0.3) + (180:0.4) $) -- ($ (0,-0.3) + (-170:0.7) $) node[label=(180+\cmdKV@general@rotate):\cmdKV@gLabels@eLC] {};
    \draw[\cmdKV@gTypes@eD] ($ (0,0.3) + (180:0.4) $) -- ($ (0,0.3) + (170:0.7) $) node[label=(180+\cmdKV@general@rotate):\cmdKV@gLabels@eLD] {};
    \draw[label distance=1,\cmdKV@gTypes@iA] ($ (0,-0.3) + (0:0.4) $) -- node[label=(-90+\cmdKV@general@rotate):\cmdKV@gLabels@iLA] {} ($ (1,0) + (-135:0.26) $);
    \draw[label distance=1,\cmdKV@gTypes@iB] ($ (0,-0.3) + (135:0.26) $) -- node[label=(-180+\cmdKV@general@rotate):\cmdKV@gLabels@iLB] {}($ (0,0.3) + (-135:0.26) $);
    \draw[label distance=1,\cmdKV@gTypes@iC] ($ (0,0.3) + (0:0.4) $) -- node[label=(90+\cmdKV@general@rotate):\cmdKV@gLabels@iLC] {} ($ (1,0) + (135:0.26) $);
    \draw[label distance=1,\cmdKV@gTypes@iD] ($ (0,0.3) + (-45:0.26) $) -- node[label=(\cmdKV@general@rotate):\cmdKV@gLabels@iLD] {}($ (0,-0.3) + (45:0.26) $);
    \node[label=(\cmdKV@general@rotate):\cmdKV@gLabels@iLE] at (1.2,0) {};
    \node[label=(-90+\cmdKV@general@rotate):\cmdKV@gLabels@iLF] at (0,-0.5) {};
    \node[label=(90+\cmdKV@general@rotate):\cmdKV@gLabels@iLG] at (0,0.5) {};
    #2
  \end{tikzpicture}
  \gSetStdTypes{gluon}
}

\makeatother

\DeclareFontFamily{OT1}{pzc}{}
\DeclareFontShape{OT1}{pzc}{m}{it}{<-> s * [1.350] pzcmi7t}{}
\DeclareMathAlphabet{\mathpzc}{OT1}{pzc}{m}{it}

\def\cM{\mathcal{M}}
\def\cO{\mathcal{O}}
\def\cN{\mathcal{N}}

\def\cR{\mathcal{R}}

\def\d{\mathrm{d}}
\def\eps{\epsilon}

\def\Tr{\operatorname*{Tr}}

\def\trm{\operatorname*{tr}\,\!\!_{-}}
\def\trpm{\operatorname*{tr}\,\!\!_{\pm}}

\def\braket#1{\langle #1 \rangle}

\def\nn{\nonumber}

\newcommand{\widebar}{\overline}

\begin{document}
\tikzset{external/force remake}

\preprint{CERN-TH-2019-042}
\preprint{CP3-19-15}
\preprint{UUITP-14/19}
\preprint{NORDITA 2019-034}

\title{The Full-Color Two-Loop Four-Gluon Amplitude in \texorpdfstring{$\cN=2$}{N=2} Super-QCD}

\author{Claude Duhr}
\affiliation{Theoretical Physics Department, CERN,
CH-1211 Geneva 23, Switzerland.}
\affiliation{Center for Cosmology, Particle Physics and Phenomenology (CP3),
UCLouvain, Chemin du Cyclotron 2, 1348 Louvain-La-Neuve, Belgium.}
\author{Henrik Johansson}
\affiliation{Department of Physics and Astronomy, Uppsala University,
75108 Uppsala, Sweden}
\affiliation{Nordita, Stockholm University and KTH Royal Institute of Technology,\\
Roslagstullsbacken 23, 10691 Stockholm, Sweden}
\author{Gregor K\"{a}lin}
\author{Gustav Mogull}
\affiliation{Department of Physics and Astronomy, Uppsala University, 75108 Uppsala, Sweden}
\author{Bram Verbeek}
\affiliation{Center for Cosmology, Particle Physics and Phenomenology (CP3),
UCLouvain, Chemin du Cyclotron 2, 1348 Louvain-La-Neuve, Belgium.}

\begin{abstract}
We present the fully integrated form of the two-loop four-gluon amplitude in $\cN=2$ supersymmetric quantum chromodynamics
with gauge group SU$(N_c)$ and with $N_f$ massless supersymmetric quarks (hypermultiplets) in the fundamental representation.
Our result maintains full dependence on $N_c$ and $N_f$, and relies on the existence of a compact integrand representation that exhibits the duality between color and kinematics.
Specializing to the $\cN=2$ superconformal theory, where $N_f=2N_c$, we obtain remarkably simple amplitudes that have an analytic structure close to that of $\cN=4$ super-Yang-Mills theory, except that now certain lower-weight terms appear. We comment on the corresponding results for other gauge groups. 
\end{abstract}

\maketitle


The formidable goal of one day solving a four-dimensional gauge theory such as quantum chromodynamics (QCD) has inspired spectacular progress related to analytic computations of scattering amplitudes. Most of the progress targets the simpler $\cN=4$ Super Yang-Mills (SYM) theory, where analytic results for multi-leg amplitudes are known to very high loop orders~\cite{Anastasiou:2003kj,*[{ \\}][{}]Bern:2005iz,DelDuca:2009au,*DelDuca:2010zg,Goncharov:2010jf,Dixon:2011pw,*Dixon:2011nj,Dixon:2013eka,*[{ \\}][{}]Dixon:2014iba,Dixon:2014voa,*[{ \\}][{}]Drummond:2014ffa,Dixon:2015iva,Dixon:2016nkn,*[{ \\}][{}]Caron-Huot:2016owq,Drummond:2018caf,Caron-Huot:2019vjl,Golden:2014xqa,*[{ \\}][{}]Harrington:2015bdt,*[{ \\}][{}]Golden:2014pua,Bern:1997nh,Naculich:2008ys,Henn:2016jdu,Abreu:2018aqd,Chicherin:2018yne}.

Massless higher-loop amplitudes often evaluate to multiple polylogarithms (MPLs)~\cite{Goncharov:1998kja,*GoncharovMixedTate}, which generalize the ordinary logarithm and dilogarithms functions. The notion of a \emph{transcendental weight}, which counts the number of integrations in the functions' definition, has played an essential role for classifying the MPLs that are allowed to appear.
The logarithm (and $i\pi=\log(-1)$) has weight one, while the dilogarithm has weight two, {\it etc}.
It was empirically observed in many examples that an $L$-loop amplitude in $\cN=4$ SYM always has uniform weight~$2L$.
The uniform weight property of $\cN=4$ SYM was not only observed for scattering amplitudes, but it was also established for certain anomalous dimensions~\cite{Kotikov:2001sc,*Kotikov:2002ab,Kotikov:2004er,Kotikov:2007cy,Grozin:2015kna,Beisert:2006ez,Almelid:2015jia,Dixon:2017nat}, form factors~\cite{vanNeerven:1985ja,Gehrmann:2011xn,*[{ \\}][{}]Brandhuber:2012vm,Brandhuber:2017bkg,*[{ \\}][{}]Brandhuber:2014ica}, correlation functions~\cite{Eden:1998hh,*[{ \\}][{}]GonzalezRey:1998tk,*[{ \\}][{}]Bianchi:2000hn,Drummond:2013nda}, and correlators of semi-infinite Wilson lines~\cite{Li:2014bfa,*Li:2014afw}.

The conjectured uniform weight property has made it possible to circumvent explicit loop calculations and instead bootstrap the function space and determine the amplitude from knowledge of kinematic limits~\cite{Dixon:2011pw,Dixon:2011nj,Dixon:2013eka,Dixon:2014iba,Dixon:2014voa,Drummond:2014ffa,Dixon:2015iva,Dixon:2016nkn,Caron-Huot:2016owq,Drummond:2018caf,Caron-Huot:2019vjl,Almelid:2017qju}.
Understanding the origin of this property, and the theories to which it can be applied, is  central for unraveling the mathematical structure of more general gauge theories. 
As of yet, there is no clear picture of which theories have amplitudes of uniform weight, or how deviations from it can be best understood. Amplitudes in QCD do not have uniform weight, though there is accumulating evidence that amplitudes in $\cN=8$ supergravity have the same uniform weight as in $\cN=4$ SYM~\cite{Brandhuber:2008tf,*[{ \\}][{}]Naculich:2008ew,*[{ \\}][{}]Chicherin:2019xeg,*[{ \\}][{}]Abreu:2019rpt,*[{ \\}][{}]Henn:2019rgj}. 

More general understanding comes from studying the BFKL gluon Green's function to high loop orders in SU($N_c$) Yang-Mills theory with generic matter content~\cite{DelDuca:2017peo}. It was observed that a necessary condition for obtaining results with uniform weight is that the matter content coincides with that of a weakly-coupled superconformal gauge theory, such as $\cN=4$ SYM, or corresponding superconformal $\cN=2,1$ theories. However, it is not expected to be a sufficient condition, and additional data concerning the weight properties of more general gauge theories is needed. Furthermore, if weight properties are to have lasting impact on bootstrapping techniques for the real-world problem of QCD, one needs to develop insight for better controlling the deviations from uniform weight.

In this Letter we study a two-loop amplitude in SU($N_c$) $\cN=2$ supersymmetric QCD (SQCD) ---
a theory which has tuneable matter content like QCD,
namely $N_f$ supersymmetric quarks, as well as a weakly-coupled superconformal phase,
like $\cN=4$ SYM, at the critical point $N_f=2N_c$.
The Lagrangian of $\cN=2$ SQCD can be constructed as the unique
$\cN=2$ supersymmetric extension of QCD.
Its perturbative spectrum consists of an adjoint vector multiplet containing the gluon field,
two gluinos, and a complex scalar.
Matter fields assemble into $N_f$ fundamental hypermultiplets, each containing a quark and two complex scalars.
We only consider the limit where the quarks and superpartners are massless.

\renewcommand\thesubfigure{\alph{subfigure}}
\begin{figure*}
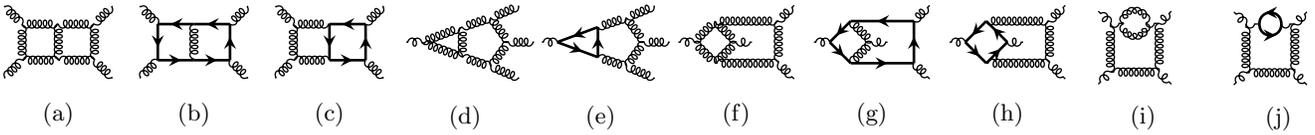

  \centering
  \begin{subfigure}{.095\textwidth}
    \centering
    \gBoxBox[scale=.8]{}
    \caption{}
  \end{subfigure}
  \begin{subfigure}{.095\textwidth}
    \centering
    \gBoxBox[scale=.8,iA=aquark,iB=aquark,iF=aquark,iE=aquark,iD=aquark,iC=aquark]{}
    \caption{}
  \end{subfigure}
  \begin{subfigure}{.095\textwidth}
    \centering
    \gBoxBox[scale=.8,iA=aquark,iB=aquark,iF=aquark,iG=quark]{}
    \caption{}
  \end{subfigure}
  \begin{subfigure}{.095\textwidth}
    \centering
    \gTriPenta[scale=.68]{}
    \caption{}
  \end{subfigure}
  \begin{subfigure}{.095\textwidth}
    \centering
    \gTriPenta[scale=.68,iD=aquark,iE=aquark,iG=aquark]{}
    \caption{}
  \end{subfigure}
  \begin{subfigure}{.095\textwidth}
    \centering
    \gBoxBoxNP[scale=.7]{}
    \caption{}
  \end{subfigure}
  \begin{subfigure}{.095\textwidth}
    \centering
    \gBoxBoxNP[scale=.7,iA=aquark,iB=aquark,iC=aquark,iD=aquark,iE=aquark]{}
    \caption{}
  \end{subfigure}
  \begin{subfigure}{.095\textwidth}
    \centering
    \gBoxBoxNP[scale=.7,iC=aquark,iD=aquark,iF=aquark,iG=aquark]{}
    \caption{}
  \end{subfigure}
  \begin{subfigure}{.095\textwidth}
    \centering
    \gBoxBubA[scale=.81,iE=agluon,iF=agluon]{}
    \caption{}
  \end{subfigure}
  \begin{subfigure}{.095\textwidth}
    \centering
    \gBoxBubA[scale=.81,iE=aquark,iF=aquark]{}
    \caption{}
  \end{subfigure}
  \caption{\small Ten cubic diagrams that describe the four-gluon two-loop amplitude of $\cN=2$ SQCD. At the conformal point, $N_f=2N_c$, diagrams (d), (e), (i), and (j) manifestly cancel out from the integrand.
    \label{fig:masterdiags}}
\end{figure*}

As the main result, we present a closed analytic form of the two-loop four-gluon amplitude in $\cN=2$ SQCD, for arbitrary $N_c$ and $N_f$.
Our starting point is the integrand of Ref.~\cite{Johansson:2017bfl},
which is characterized by a particularly elegant presentation that exhibits
the duality between color and kinematics~\cite{Bern:2008qj,*Bern:2010ue,Johansson:2015oia}.
The duality was critical for obtaining the non-planar integrand contributions
through kinematic identities that relate them to the simpler planar ones.
It was later observed that the numerators display a remarkable simplicity when re-expressed in terms of Dirac traces~\cite{Kalin:2018thp},  from which one may expect that the final result should share some of the simplicity of the integrand.

\section{The Color-dual integrand}
$\cN=2$ SQCD has a running coupling constant $\alpha_S(\mu_R^2)$,
and loop amplitudes need to be renormalized to remove ultraviolet (UV) divergences.
Prior to renormalization,
we may perturbatively expand $n$-point amplitudes in terms of the bare coupling $\alpha_S^0$,
\begin{equation}\label{eq:fullAmplitude}
\cM_n=(4\pi\alpha_S^0)^{\frac{n-2}2}
\sum_{L=0}^\infty\left(\frac{\alpha^0_SS_\eps}{4\pi}\right)^L\cM_n^{(L)}\,,
\end{equation}
where $S_{\epsilon} = (4\pi)^{\epsilon}e^{-\epsilon \gamma}$
anticipates dimensional regularization in $D=4-2\epsilon$ dimensions.
At multiplicity four, we will always use kinematics defined by  $s>0$; $t,u<0$, where $s=(p_1+p_2)^2$, $t=(p_2+p_3)^2$, $u=(p_1+p_3)^2$, with $s+t+u=0$, and all momenta are outgoing.

The two-loop integrand of Ref.~\cite{Johansson:2017bfl} was constructed to make manifest separations at the diagrammatic level between distinct gauge-invariant contributions. For example, it manifests the difference between $\cN=4$ SYM and the $\cN=2$ superconformal theory \mbox{(SCQCD)}, with $N_f=2N_c$, as a combination of simple diagrams that are manifestly UV finite.
The diagrammatic separation will allow us to have a clear partition of the integrated answer into terms with distinct physical interpretation.

The integrand of Ref.~\cite{Johansson:2017bfl} consists of 19 cubic diagrams. 
Of these only ten, shown in Fig.~\ref{fig:masterdiags}, give rise to non-vanishing integrals.
Using these ten diagrams (a--j) the $\cN=2$ SQCD amplitude is assembled,
with an ${S}_4$ permutation sum over external particle labels, as
\begin{equation}\label{eq:colorKinDualAmp}
i\cM_4^{(2)}=
e^{2\eps\gamma}\sum_{{S}_4}\sum_{i \in \{\rm a,\ldots, j \}}
\int\!\frac{\d^{2D}\ell}{(i\pi^{D/2})^2}
\frac{(N_f)^{|i|}}{S_i}\frac{n_i c_i}{D_i}\,,
\end{equation}
where $\d^{2D}\ell\equiv\d^D\ell_1\d^D\ell_2$ is the two-loop integration measure
and $|i|$ is the number of matter loops in the given diagram.
Diagrams are described by kinematic numerator factors $n_i$, color factors $c_i$,
symmetry factors $S_i$, and propagator denominators $D_i$~\cite{Johansson:2017bfl,Kalin:2018thp}.

Color-kinematics duality requires that the kinematic numerators $n_i$ satisfy
the same general Lie algebra relations as the color factors $c_i$~\cite{Bern:2008qj,*Bern:2010ue}.
Through these relations the numerators of diagrams (e--j) were completely determined by the four planar diagrams (a--d).
The integrand was constructed through an Ansatz constrained to satisfy $(D\le 6)$-dimensional unitarity cuts. The upper bound corresponds to the $D=6$, $\cN=(1,0)$ SQCD theory, which is the unique supersymmetric maximal uplift of the four-dimensional $\cN=2$ SQCD  theory.

For later convenience we quote the relevant numerator contributions to diagram (b) for different gluon helicities,
\begin{subequations}\label{eq:doubleBoxes}
\begin{align}
\label{eq:doubleBox12}
n\!\left(\!\gBoxBox[scale=.6,eLA=$1^-$,eLB=$2^-$,eLC=$3^+$,eLD=$4^+$,
iA=aquark,iB=aquark,iF=aquark,iE=aquark,iD=aquark,iC=aquark,
iLA=$\downarrow\!\ell_1$,iLD=$\ell_2\!\downarrow$]{}\!\!\right)
&=-\kappa_{12} \mu_{12}  \,,\\
\label{eq:doubleBox13}
n\!\left(\!\gBoxBox[scale=.6,eLA=$1^-$,eLB=$2^+$,eLC=$3^-$,eLD=$4^+$,
iA=aquark,iB=aquark,iF=aquark,iE=aquark,iD=aquark,iC=aquark,
iLA=$\downarrow\!\ell_1$,iLD=$\ell_2\!\downarrow$]{}\!\!\right)
&=\frac{ \kappa_{13} }{u^2}\trm(1\ell_124\ell_23)\,,\\
\label{eq:doubleBox14}
n\!\left(\!\gBoxBox[scale=.6,eLA=$1^-$,eLB=$2^+$,eLC=$3^+$,eLD=$4^-$,
iA=aquark,iB=aquark,iF=aquark,iE=aquark,iD=aquark,iC=aquark,
iLA=$\downarrow\!\ell_1$,iLD=$\ell_2\!\downarrow$]{}\!\!\right)
&=\frac{\kappa_{14}}{t^2}\trm(1\ell_123\ell_24) \,,
\end{align}
\end{subequations}
where $\kappa_{ij}$ is proportional to the color-stripped tree amplitude ---
for instance, in the purely gluonic case we have $\kappa_{12}=ist \,M^{(0)}_{(--++)}$,
and $M^{(0)}_{(--++)} = -i\braket{12}^2[34]^2/st$.
The $\trpm$ are chirally projected Dirac traces taken strictly over the four-dimensional parts of the momenta, and $\mu_{ij}=-\ell_i^{[-2\eps]}\cdot\ell_j^{[-2\eps]}$ contain the extra-dimensional loop momenta.
Numerators of the other diagrams are of comparable simplicity, see Refs.~\cite{Johansson:2017bfl,Kalin:2018thp}.
Note that four-gluon amplitudes with helicity $(\pm{+}{+}{+})$ exactly vanish in supersymmetric theories due to Ward identities, so we need not consider them. Without loss of generality we focus on gluon amplitudes with helicity configurations $({-}{-}{+}{+})$ and $({-}{+}{-}{+})$; the general vector multiplet cases are obtained from these by supersymmetric Ward identities. 

In general, we note that any gluon amplitude in $\cN=2$ SQCD can be decomposed into
three independent blocks that have different characteristics after integration,
\begin{equation}\label{eq:MRS}
\cM_n^{(L)}=\cM_n^{(L)[\cN=4]} +\cR_n^{(L)} + (C_A-N_f) {\cal  S}_n^{(L)}\,.
\end{equation}
$C_A$ is the quadratic Casimir; for SU($N_c$) we normalize it as $C_A=2N_c$. The  decomposition involves three terms: $\cM_n^{(L)[\cN=4]}$ is a gluon amplitude in $\cN=4$ SYM, $\cR_n^{(L)}$ is a remainder function that survives at the conformal point  $N_f=2N_c$,
and ${\cal  S}_n^{(L)}$ is a term that contributes away from the conformal point.
By definition the first term will have the same
(uniform) weight property as $\cN=4$ SYM,
and the first two terms will be free of UV divergences. The two-loop integrand of Eq.~(\ref{eq:colorKinDualAmp}) is particularly well suited to this decomposition:
only four diagrams (b,c,g,h) contribute to $\cR_4^{(2)}$,
and six diagrams (b,c,e,g,h,j) contribute to ${\cal  S}_4^{(2)}$.
Diagrams (a,d,f,i) may be ignored as they only contribute to the two-loop $\cN=4$ SYM amplitude,
which is already known in the literature~\cite{Bern:1997nh,Anastasiou:2003kj,Naculich:2008ys}.

\section{Integrating the two-loop amplitude}

In order to promote the integrand (\ref{eq:colorKinDualAmp})
to a fully integrated two-loop amplitude, the following steps are taken.
First, all contributions are reduced to scalar-type integrals in shifted dimensions 
using Schwinger parametrization (see e.g.~Ref.~\cite{Bern:2002tk}).
This technique works particularly well when Dirac traces are involved as the number of resulting scalar integrals tends to be small.
The extra-dimensional components $\mu_{ij}$ are treated in the same way.

Next, the higher-dimensional scalar integrals are reduced to $D$ dimensions using dimensional recurrence relations~\cite{Anastasiou:2003kj,Tarasov:1996br,*[{ \\}][{}]Smirnov:1999wz,*[{ \\}][{}]Lee:2009dh,Anastasiou:2000mf,*[{ \\}][{}]Anastasiou:2000kp}.
The resulting integrals are reduced to a basis of master integrals using integration-by-parts (IBP) relations~\cite{Chetyrkin:1981qh,*[{ \\}][{}]Tkachov:1981wb}, for which we used the \emph{Mathematica} package LiteRed~\cite{Lee:2012cn}.
These master integrals are known analytically~\cite{Tausk:1999vh,Anastasiou:2000mf,Gehrmann:2005pd,Henn:2013pwa}, and their insertion yields our final result. Manipulation of the master integrals expressed in terms of harmonic polylogarithms~\cite{Remiddi:1999ew} was done using the \emph{Mathematica} package HPL~\cite{Maitre:2005uu}.

The complete one- and two-loop amplitudes are presented in {\it Mathematica}-readable ancillary text files~\cite{ancillary_file}. (The one-loop results are adapted from Ref.~\cite{Johansson:2014zca})
We have performed several checks on our result.
First, we have integrated an alternative representation of the integrand ---
given also in Ref.~\cite{Johansson:2017bfl} ---
non-trivially obtaining the same result.
Second, we have checked that our result reproduces the high-energy behavior expected from the
known two-loop Regge trajectory for supersymmetric gauge theories \cite{Kotikov:2000pm}.
Finally, the amplitude is divergent,
and we have checked that the amplitude has the correct IR-pole structure after UV renormalization.

UV divergences are captured by the $\beta$-function. From the all-order NSVZ $\beta$-function~\cite{Novikov:1983uc} it can be seen that the $\beta$-function of $\cN=2$ SQCD is one-loop exact,
\begin{equation}\label{eq:betaFunction}
\beta(\alpha_S(\mu_R^2))=
-\alpha_S(\mu_R^2)\,\left(2\eps+\beta_0\frac{\alpha_S(\mu_R^2)}{2\pi}\right)\,,
\end{equation}
where $\beta_0=C_A-T_RN_f$, with $C_A=2N_c$ and we take $T_R=1$ in this Letter. 
UV-renormalized amplitudes $\widetilde{\cM}_n^{(L)}$ are defined by Eq.~(\ref{eq:fullAmplitude}),
except with $\alpha_S^0S_\eps$ replaced by $\alpha_S(\mu_R)$.
In the $\widebar{\text{MS}}$ scheme the renormalized and bare couplings are related by
\begin{equation}
\alpha^0_SS_{\epsilon}=
\alpha_S(\mu_R^2) \mu_R^{2\epsilon}
\sum_{L=0}^\infty\left(-\frac{\beta_0}{\eps}\frac{\alpha_S(\mu_R^2)}{4\pi}\right)^L\,.
\end{equation}
See e.g.~Ref.~\cite{Catani:1998bh} for more details.

The IR singularities of a scattering amplitude are universal and independent of the hard scattering process.
At one loop a color-space operator $\mathbf{I}^{(1)}(\epsilon)$ may be defined
that encodes all IR singularities~\cite{Catani:1998bh},
\begin{equation}\label{eq:Catani_1}
\widetilde{\cM}_n^{(1)}=
\widetilde{\cM}_n^{(1)\text{fin}}+\mathbf{I}^{(1)}(\epsilon)\widetilde{\cM}_n^{(0)}\,,
\end{equation}
where $\widetilde{\cM}_n^{(L)\text{fin}}$ is finite as $\eps\to0$.

The IR singularities of two-loop amplitudes are encoded into the formula
\cite{Catani:1998bh,Sterman:2002qn,*Becher:2009cu,*Gardi:2009qi}
\begin{align}\nonumber
&\widetilde{\cM}_n^{(2)} = \widetilde{\cM}_n^{(2)\text{fin}} + \mathbf{I}^{(1)}(\epsilon)\widetilde{\cM}_n^{(1)}
-\bigg[\frac12\mathbf{I}^{(1)}(\epsilon)^2+\frac{\beta_0}{\epsilon}\mathbf{I}^{(1)}(\epsilon)\\
\label{eq:Catani_2}&\!-e^{-\epsilon \gamma}\!\frac{\Gamma(1-2\epsilon)}{\Gamma(1-\epsilon)}\!\left(\! \frac{\beta_0}{\epsilon} + K \!\right)\mathbf{I}^{(1)}(2 \epsilon)- \mathbf{H}^{(2)}(\epsilon)\! \bigg]\cM_n^{(0)}.
\end{align}
In our case we have $K=-\zeta_2 \, C_A + 2(1+2\eps)\beta_0$.
The tensor $\mathbf{H}^{(2)}(\epsilon)$,
which determines the $\mathcal{O}(\epsilon^{-1})$ pole, is
\begin{equation}
  \mathbf{H}^{(2)}(\epsilon) = \frac{e^{\epsilon\gamma}}{4 \epsilon \Gamma(1-\epsilon)}\left( \frac{\mu_R^2}{-s} \right)^{2 \epsilon} \left( 4 H_g^{(2)} \mathbf{1} + \mathbf{\widehat{H}}^{(2)}\ \right)\,.
\end{equation}
The color-diagonal part is fixed by
$H_g^{(2)}=C_A^2\zeta_3/2+\beta_0\left(2\beta_0+C_A\zeta_2/4\right)$.
The non-diagonal part $\mathbf{\widehat{H}}^{(2)}$ matches what
was found in Ref.~\cite{Bern:2002tk} for QCD and $\cN=1$ SYM.

As previously mentioned, the integrated amplitude can be decomposed as in Eq.~(\ref{eq:MRS}), where the remainder $\cR^{(2)}_4$ is strikingly compact and discussed further in the next section. The contribution ${\cal  S}_4^{(2)}$ is given in the ancillary files for the unrenormalized amplitude~\cite{ancillary_file}. We note that various SU($N_c$) color relations allow us to eliminate certain partial amplitudes in favor of the independent components.  For example, in ${\cal  S}_4^{(2)}$ only the planar leading-color part is independent under the SU($N_c$) color algebra. Our results make this property manifest~\cite{ancillary_file}. 

Having at our disposal the full analytic result for the two-loop $\cN=2$ SQCD amplitude, we can use it to study its transcendental weight properties as a function of $N_f$ and $N_c$. This is the first time that the weight properties of a full-color two-loop amplitude are explored in a gauge theory which is neither superconformal nor equal to QCD (or the pure Yang-Mills theory). We observe that the unrenormalized amplitude in $\cN=2$ SQCD for generic $N_f$ and $N_c$ shares the same weight properties as the two-loop QCD amplitude. In particular, the coefficient of $\eps^{-k}$ of the function ${\cal  S}_4^{(2)}$ contains MPLs of weight up to $4-k$, including weight 0 (note that the smallest value of $k$ is $k=3$, and the coefficient of $1/\eps^3$ has weight 0). 

\section{The remainder of SCQCD}

Consider the UV-finite remainder function $\cR^{(L)}_n$ in Eq.~\eqref{eq:MRS}
needed to describe the $\cN=2$ SCQCD amplitude.
By definition $\cR^{(L)}_n$  vanishes at tree level,
and it also vanishes at one loop in the large-$N_c$ limit for any number of external legs
\cite{Glover:2008tu}.
Beyond leading color, $\cR_n^{(1)}$ is non-zero and IR finite.
At four points, we read off the coefficient of
$ M^{(0)}_{(--++)}N_c^0\Tr(T^{a_1}T^{a_2})\Tr(T^{a_3}T^{a_4})$,
which we denote by  $R_{(--)(++)}^{(1)[0]}$ (and analogously for other helicity configurations),
giving \cite{Johansson:2014zca}
\begin{subequations}
\begin{align}
R_{(--)(++)}^{(1)[0]}&=
2\tau\left[ (T-U)^2 + 6\zeta_2 \right]+\cO(\eps)\,,\\
R_{(-+)(-+)}^{(1)[0]}&=
\frac{2\tau}{\upsilon^2} T(T+2 i \pi) +\cO(\eps)\,,
\end{align}
\end{subequations}
where we have introduced the shorthand notation $\tau = -t/s$, $\upsilon = -u/s$,
$T = \log(\tau)$, and $U = \log(\upsilon)$.
We recall that $s>0$; $t,u<0$, so $T$ and $U$ are real.
We see that the one-loop remainders have uniform transcendental weight two, just like the corresponding $\cN=4$ SYM amplitudes.

The two-loop remainder functions are finite and non-zero already at leading color~\cite{Dixon2008talk,Leoni:2015zxa}.
Beyond leading color, the two-loop remainders develop IR divergences which can be cast in a form reminiscent of Eq.~\eqref{eq:Catani_1},
\begin{equation}\label{eq:RemIRSub}
\cR_n^{(2)}=\cR_n^{(2)\text{fin}}+\mathbf{I}^{(1)}(\epsilon)\cR_n^{(1)}\,,
\end{equation}
with $\cR_n^{(2)\text{fin}}$ finite as $\eps\to0$.
Using the integrand in Eq.~(\ref{eq:colorKinDualAmp}) we can easily isolate individual diagrams that contribute to $\cR_n^{(2)}$ at different orders of $N_c$.
For example, at leading color ${\cal O}(N_c^2)$, only graph (b) is needed to compute the SCQCD remainder. The coefficient of $N_c^2\Tr(T^{a_1}T^{a_2}T^{a_3}T^{a_4})$ can be compactly written as
\begin{align}\label{eq:LCRemDiag}
\!\!R_{(1234)}^{(2)[2]}
=\frac{-i}{\pi^D}e^{2\eps\gamma}\!\sum_\text{cyclic}
\int\!\frac{\d^{2D}\ell}{D_\text{b}}
n\!\left(\!\!\gBoxBox[scale=.6,eLA=$1$,eLB=$2$,eLC=$3$,eLD=$4$,
iA=aquark,iB=aquark,iF=aquark,iE=aquark,iD=aquark,iC=aquark,
iLA=$\downarrow\!\ell_1$,iLD=$\ell_2\!\downarrow$]{}\!\!\right)\,,
\end{align}
where the numerators are given in Eq.~(\ref{eq:doubleBoxes}), and $D_\text{b}$ contains the propagator denominators of diagram (b).

As the integral with the $\mu_{12}$ factor in
Eq.~(\ref{eq:doubleBox12}) begins at $\cO(\eps)$, we need only integrate the two Dirac traces in
Eqs.~(\ref{eq:doubleBox13}) and (\ref{eq:doubleBox14}).
They are manifestly IR finite, giving the leading-color remainders (for $s>0$; $t,u<0$)
\begin{subequations}\label{eq:LCReminder2}
\begin{align}
&R^{(2)[2]}_{(--++)}=12\zeta_3 + \frac{\tau}{6}\big\{48 \text{Li}_4(\tau )-24 T\text{Li}_3(\tau )\nn\\
&-24 T \text{Li}_3(\upsilon )+24\text{Li}_2(\tau ) \left(\zeta _2+T U\right)+24 T U\text{Li}_2(\upsilon )\nn\\
&-24 U \text{Li}_3(\tau )-24S_{2,2}(\tau )+T^4-4 T^3 U+18 T^2 U^2\nn\\
&-12 \zeta _2 T^2+24 \zeta _2 T U+24 \zeta _3 U-168 \zeta _4-4 i \pi  \big[6 \text{Li}_3(\tau )\nn\\
&+6 \text{Li}_3(\upsilon )-6 U \text{Li}_2(\tau )-6 U \text{Li}_2(\upsilon )-T^3+3T^2 U\nn\\
\label{eq:LC_R--++}&-6 T U^2-6 \zeta _2 T+6 \zeta _2 U\big]\big\}+\cO(\eps)\,,\\
\label{eq:LC_R-+-+}&R^{(2)[2]}_{(-+-+)}=12\zeta_3+\frac1{6}\frac{\tau}{\upsilon^2}T^2(T+2 i \pi)^2+\cO(\eps)\,,
\end{align}
\end{subequations}
where $\text{Li}_n(z)$ are the classical polylogarithms
and $S_{n,p}(z)$ are Nielsen generalized polylogarithms.
$R^{(2)[2]}_{(-+-+)}$ is comparatively simpler as cancellations occur between
two cyclic permutations of the numerator~(\ref{eq:doubleBox13}). The leading-color remainder of $\cN=2$ SCQCD was presented in Ref.~\cite{Dixon2008talk}, and the $R^{(2)[2]}_{(- -++)}$ remainder was first published in Ref.~\cite{Leoni:2015zxa};  we confirm the correctness of those results. Furthermore, the diagrams in (\ref{eq:LCRemDiag}) precisely match the simple IR finite integrals considered in Ref.~\cite{CaronHuot:2012ab}.

Inspecting the leading-color result, it is clear that remainder does not have uniform weight four, in agreement with Refs.~\cite{Dixon2008talk,Leoni:2015zxa}. The deviation from uniform weight, however, is very minimal and entirely captured by a constant $12\zeta_3$ times the tree amplitude. This 
leads to some hope that the deviation from maximal weight is simple enough that it can be understood in general, at least at leading color. 

The subleading-color part of $\cR_4^{(2)}$ is
given by the coefficients of $ M^{(0)}_{(--++)} N_c\Tr(T^{a_1}T^{a_2})\Tr(T^{a_3}T^{a_4})$.
We find for the finite parts
\begin{subequations}
\begin{align}
&R^{(2)[1]\text{fin}}_{(--)(++)}=  \frac{2\tau}{3} \big\{96 \text{Li}_4(\tau )-72 T \text{Li}_3(\tau )+24 T\text{Li}_3(\upsilon )\nn\\
&+24 T \text{Li}_2(\tau )
   (T-U)-24 U \text{Li}_2(\upsilon ) (T-U)+96
   \text{Li}_4(\upsilon )\nn\\
&+24 U \text{Li}_3(\tau )-72 U
   \text{Li}_3(\upsilon )+T^4+4 T^3 U-18
   T^2 U^2\nn\\
&+4 T U^3+U^4+24 \zeta _2 T U-12 \zeta _2 T^2-12 \zeta _2 U^2-654 \zeta _4\nn\\
&-4 i \pi  \big[12
   \text{Li}_3(\tau )+12
   \text{Li}_3(\upsilon )-12 T \text{Li}_2(\tau )-12 U \text{Li}_2(\upsilon
   )\nn\\
&-T^3-3 T^2 U-3 T U^2-U^3-18 \zeta _2 T-18 \zeta _2
   U\big]\big\} \nn \\ & +\cO(\eps)\,,\label{eq:NLCRemainder1}\\
&R^{(2)[1]\text{fin}}_{(-+)(-+)}=\frac{2\tau}{3\upsilon^2}\big\{ 48 \text{Li}_4(\tau )-24 T \text{Li}_3(\tau )-24 S_{2,2}(\tau )\nn\\
&+24 \zeta _2 \text{Li}_2(\tau )+T^4-84 \zeta _2 T^2-102 \zeta _4+24 T \zeta_3 \nn\\
&-8 i \pi \big[3 T \zeta_2- T^3\big] \big\} - \frac{8 \tau}{3 \upsilon^2}\big\{ 6 \tau  \text{Li}_3(\tau )-6 \tau  \text{Li}_3(\upsilon )\nn\\
&-6 \tau  T \text{Li}_2(\tau )+6\text{Li}_3(\upsilon )-6 \upsilon U \text{Li}_2(\upsilon )+3 \tau  T U^2\nn\\
&+3 T \upsilon  U^2 -3 T U^2 -30 \tau  T \zeta _2-30 \upsilon  U  \zeta _2-6 \zeta_3\nn\\
&+3 i \pi  \big[2(\upsilon-\tau ) \text{Li}_2(\tau )+\tau  T^2 +2 T \upsilon  U+\upsilon  U^2 + 2\tau \zeta_2\big]  \big\}\nn \\
& +\cO(\eps) \,. \label{eq:NLCRemainder2}
\end{align}
\end{subequations}
Finally, the subsubleading-color parts of $\cR_4^{(2)}$ are,
for all helicity configurations, given by the relation
\begin{align} \label{eq:NLCReminder2}
R^{(2)[0]}_{(1234)}&=R^{(2)[2]}_{(1234)}-R^{(2)[1]}_{(13)(24)}+\frac12\left(R^{(2)[1]}_{(12)(34)}+R^{(2)[1]}_{(14)(23)}\right)\!.
\end{align}
It follows from consistency conditions on the SU($N_c$) color algebra,
and using the fact that only four diagrams (b,c,g,h) 
are required to fully specify $\cR_4^{(2)}$.  We have explicitly checked that this relation is satisfied by the integrated amplitude.

Equations (\ref{eq:RemIRSub}) and (\ref{eq:LCReminder2})--(\ref{eq:NLCReminder2}) give the full analytic result for the remainder $\cR_4^{(2)}$.
We find it striking that the complete result for four-gluon scattering at two loops in $\cN=2$
SCQCD can be cast in a compact form which fits into a few lines.
Analyzing the weight properties of $\cR_4^{(2)}$,
we observe that the finite terms of the subleading-color part involve MPLs of weight 2,
3 and 4, but no lower-weight MPLs are present.
Moreover, we observe a striking cancellation between lower-weight terms in the
two-loop remainder and the higher-order terms in $\eps$ in Eq.~\eqref{eq:RemIRSub}.
As a result, the finite remainders $\cR_4^{(2)\text{fin}}$ in
Eqs.~\eqref{eq:NLCRemainder1} and~\eqref{eq:NLCRemainder2} only involve MPLs of weight 3 and 4.
In other words, there seems to be conspiracy between the lower-weight terms and
the structure of the IR divergences described by Eq.~\eqref{eq:RemIRSub},
resulting in a minimal departure from the property of maximal weight in the finite remainder.
It would be interesting to understand better this interplay between the structure of infrared
divergences and the appearance of lower-weight terms.

Finally, we note that the integrand (\ref{eq:colorKinDualAmp}) is valid for any gauge group and matter representation --- for convenience, results for U($N_c$), SO($N_c$), and USp($N_c$) groups are included in the ancillary files~\cite{ancillary_file} --- making it possible to study the group-theory impact on the weight properties. For abelian U(1) $\cN=2$ SQED, only diagram (b) is non-vanishing; hence the full amplitude is obtained by summing over all permutations of the external legs in the leading-color remainders Eqs.~\eqref{eq:LC_R--++} and \eqref{eq:LC_R-+-+},
multiplied by the respective tree amplitudes.
Since the weight-3 terms are constants,
they cancel in the sum due to a photon-decoupling identity
satisfied by the tree amplitudes.
As a result, the two-loop $\cN=2$ SQED amplitude has uniform weight four,
in agreement with Ref.~\cite{Binoth:2002xg}.

A curious observation is the following: if we work with gauge group SO($3$) and a fundamental hypermultiplet ($N_f=1$), the amplitude in $\cN=2$ SQCD is identical to the amplitude in SO($3$) $\cN=4$ SYM, and thus has uniform transcendental weight. We expect this equality to hold to any loop order since both the structure constants and fundamental generators of SO(3) are described by rank-3 Levi-Civita tensors, so the fundamental hypermutiplet behaves as if it belonged to the adjoint representation. This observation shows that the weight properties are tightly connected not only to the matter content and the symmetries of the theory, but also the choice of gauge group and representation play an important role.

\section{Conclusions and outlook}

In this Letter we have presented the fully integrated two-loop four-gluon $\cN=2$ SQCD amplitude,
generalizing previous results to full $N_c$ and $N_f$ dependence.
Using our result we have, for the first time,
performed a complete analysis of the transcendental weight properties of an amplitude
beyond one loop and beyond leading color in a gauge theory that is neither $\cN=4$ SYM nor QCD.
While the one-loop amplitude has uniform weight two, just like in $\cN=4$ SYM,
we find that for generic matter content the two-loop amplitude contains
MPLs of all possible allowed weights, just like in QCD.

Interestingly, when restricting to the conformal point
many of the lower-weight terms in the amplitude disappear.
Moreover, we observe striking cancellations between lower-weight terms
when the infrared poles are subtracted.
Based on these observations
one may speculate that conformal symmetry and infrared singularities play a vital role in understanding the detailed transcendental weight of scattering amplitudes.
Finally, we observe that the choice of the gauge group also has an impact on
the transcendental weight,
and we have identified two instances where a gauge group
other than SU($N_c$) leads to amplitudes of uniform and maximal weight.

Looking forward, it will be interesting to extend the analysis of transcendental weight to higher loops, higher multiplicities, or to $\cN<2$ supersymmetric theories. In the latter case, further decomposition into superconformal remainders may reveal additional hidden structure.
Generalizing our $\cN=2$ SQCD amplitudes to massive hypermultiplets, or to other external state configurations would also be interesting. The latter study has been initiated in Ref.~\cite{Kalin:2018thp},
where color-kinematics-dual integrands for two-loop amplitudes with external hypermultiplets are available.

\begin{acknowledgments}
\emph{Acknowledgements:} We thank Alexander Ochirov for helpful discussions, and we are grateful to Lance Dixon for useful discussions and for pointing us to previous results on planar $\cN=2$ SCQCD~\cite{Dixon2008talk}.
The research of HJ, GK, and GM is supported by the Swedish Research Council under grant 621-2014-5722, the Knut and Alice Wallenberg Foundation under grant KAW 2013.0235, and the Ragnar S\"{o}derberg Foundation (Swedish Foundations' Starting Grant). The research of CD and BV is supported by the ERC grant 637019 ``MathAm''.
\end{acknowledgments}

\bibliography{references}

\end{document}